\documentclass[11pt]{article}
\usepackage{epsfig}
\usepackage{dcolumn}
\usepackage{bm}
\usepackage{graphicx}
\usepackage{amssymb,amsmath}
\usepackage{multirow}
\usepackage{cite,color,url}
\usepackage{xcolor}
\usepackage{adjustbox}
\usepackage{tabu}
\usepackage{textcomp}
\usepackage{array}
\usepackage[colorlinks=true,urlcolor=blue,anchorcolor=blue
,citecolor=blue,filecolor=blue,linkcolor=blue,menucolor=blue
,linktocpage=true,pdfproducer=medialab,pdfa=true]{hyperref}

\usepackage{colordvi}
\usepackage{epsfig,psfrag,rotating,soul}
\def\beqa{\begin{eqnarray}}
\def\eeqa{\end{eqnarray}}

\evensidemargin \oddsidemargin
\allowdisplaybreaks
\renewcommand{\arraystretch}{1.5}
\psfrag{lk}{$l_k$}
\psfrag{lm}{$l_m$}
\psfrag{blk}{$\bar {l}_k$}
\psfrag{blm}{$\bar {l}_m$}
\def\thefootnote{\fnsymbol{footnote}}

\let\OLDthebibliography\thebibliography
\renewcommand\thebibliography[1]{
\OLDthebibliography{#1}
\setlength{\parskip}{0pt}
\setlength{\itemsep}{0pt plus 0.3ex}}
\usepackage{makecell}

\newcommand{\eslash}{\ensuremath{{\hbox{$E$\kern-0.6em\lower-.05ex\hbox{/}\kern0.10em}}}}
\begin{document}

\thispagestyle{empty}
\begin{center}
\begin{Large}
\textbf{\textsc{A Novel Experimental Search Channel\\[0.25cm]
for Very Light Higgses in the 2HDM Type I}}
\end{Large}

\vspace{1cm}
{\sc
S.~Moretti$^{1,2}$%
\footnote{\tt \href{mailto:s.moretti@soton.ac.uk}{s.moretti@soton.ac.uk, \href{mailto:stefano.moretti@physics.uu.se}{stefano.moretti@physics.uu.se}}}, 
S.~Semlali$^{1,3}$%
\footnote{\tt
\href{mailto:souad.semlali@soton.ac.uk}{souad.semlali@soton.ac.uk}},
C.~H.~Shepherd-Themistocleous$^{3}$%
\footnote{\tt
\href{mailto:claire.shepherd@stfc.ac.uk}{claire.shepherd@stfc.ac.uk}}
}

\vspace*{.7cm}

{\sl
$^1$School of Physics and Astronomy, University of Southampton, Southampton, SO17 1BJ, UK}\\
{\sl
$^2$Department of Physics \& Astronomy, Uppsala University, Box 516, SE-751 20 Uppsala, Sweden}\\
{\sl
$^3$Particle Physics Department, Rutherford Appleton Laboratory, Chilton, Didcot, Oxon OX11 0QX, UK}

\end{center}

\vspace*{0.1cm}

\begin{abstract}
\noindent
		 We present a reinterpretation study of existing results from the CMS Collaboration, specifically, searches for light Beyond the Standard Model (BSM) Higgs pairs produced in the chain decay $pp\to H_{\rm SM}\to hh(aa)$ into a variety of final states, in the context of the CP-conserving 2-Higgs Doublet Model (2HDM) Type-I. Through this, we test the Large Hadron Collider (LHC) sensitivity to a possible new signature, $pp\to H_{\rm SM}\to ZA\to ZZ h$, with $ZZ\to jj  \mu^+\mu^-$ and $h\to b\bar b$. We perform a systematic scan over the 2HDM Type-I parameter space, by taking into account all available theoretical and experimental constraints, in order to find a region with a potentially visible signal. We investigate the significance of it through a full Monte Carlo simulation down to the {parametrised} detector level. We show that such a signal is an alternative promising channel to standard four-body searches for light BSM Higgses at the LHC already with an integrated luminosity of ${\cal L} = 300~{\rm fb}^{-1}$. For a tenfold increase of the latter, discovery should be possible over most of the allowed parameter space.
\end{abstract}
 
\def\thefootnote{\arabic{footnote}}
\setcounter{page}{0}
\setcounter{footnote}{0}
\newpage
\section{Introduction}
\label{into}

One of the main goals of the LHC machine is to investigate the individual properties (mass, width, spin, CP quantum numbers) and interactions (with both matter and forces) of the Higgs boson and to look into evidence for new physics. These  Higgs features have been probed by the ATLAS and CMS collaborations, using proton-proton ($pp$) collision data collected at centre-of-mass energies of  $\sqrt{s} = 7~\text{TeV} ,~8~\text{TeV}$ and 13 TeV for an integrated luminosity of $25~\text{fb}^{-1}$ up to $139~\text{fb}^{-1}$. Although the measurements of the Higgs mass~\cite{ATLAS:2015yey, ATLAS:2018tdk, CMS:2017dib}, spin~\cite{ATLAS:2015zhl}, width~\cite{ATLAS:2018jym,CMS:2019ekd} and couplings to SM fermions and vector bosons~\cite{ATLAS:2016neq,ATLAS:2019nkf,CMS:2018uag,ATLAS:2021vrm,ATLAS:2018ynr,ATLAS:2020qdt} are all indeed in a  good agreement with the SM theoretical predictions, the uncertainties on the SM-like Higgs couplings probed in multiple production modes for the five key decay channels $H \to \gamma \gamma\text{\cite{CMS:2021kom,ATLAS:2020pvn,CMS:2018ctp,ATLAS:2018pgp,CMS:2018gwt}},~ZZ^{*}\text{\cite{ATLAS:2018pgp,CMS:2018gwt,ATLAS:2020wny}},~WW^{*}\text{\cite{CMS:2020dvg,ATLAS:2018xbv,CMS:2018zzl, CMS:2017zyp}},~\tau \tau\text{\cite{CMS:2017zyp, ATLAS:2018ynr, CMS:2021gxc}}~{\text{and}}~b\overline{b}\text{\cite{CMS:2018gwt,ATLAS:2019yhn,ATLAS:2021qou}}$ provide signs of a possible potential BSM contributions to the total Higgs width and hints of new physics through the invisible and/or undetected decays. It is worth highlighting that indirect constraints from the current fit of couplings measurements and direct searches for $H \to inv$ (i.e., to`invisible' final states) performed by ATLAS and CMS collaborations have placed upper limits on the Branching Ratio (BR) of Higgs boson to invisible particles and undetected BSM particles at 95\% C.L. (Confidence Level) \cite{ATLAS:2019cid,ATLAS:2022yvh, CMS:2018yfx}.  

Furthermore, the Higgs self-couplings are one of the most interesting interactions that can be probed at the LHC with sufficient luminosity, although, at present (i.e., at the end of Run 2), they are not determined yet. A measurement of this interaction is one of the highest priority goals during, possibly, Run 3 and, certainly, at the High Luminosity LHC (HL-LHC), both of which would, therefore, start shedding light on the nature of the Higgs boson and the shape of the Higgs potential, which in turn has implications for the vacuum metastability, the hierarchy problem as well as the strength of the Electro-Weak (EW) phase transition. However, probing Higgs self-interactions, both trilinear and quartic couplings, in multi-Higgs production is experimentally very challenging due to the small cross section for SM-like di-Higgs production via the gluon fusion process, even at Next-to-Next-to-Leading Order (NNLO)~\cite{Grazzini:2018bsd}. Both the  ATLAS and CMS collaborations have set upper limits at 95\% C.L. on the Higgs production cross sections after performing searches in various final states, e.g.  $b\overline{b}\gamma \gamma$~\cite{CMS:2018tla,ATLAS:2021ifb}, $b\overline{b}\tau\tau$\cite{CMS:2017hea} and $b\overline{b}b\overline{b}$~\cite{CMS:2022cpr, ATLAS:2022hwc}. From the theory side, many models with an extended scalar sector can be responsible for enhanced (SM-like) di-Higgs production, like the 2-Higgs Doublet model (2HDM), the Next-to-2HDM (N2HDM) and a variety if both minimal and non-minimal Supersymmetric (SUSY) models. In fact, all such BSM scenarios also present the additional features of new di-Higgs final states, as they all present with additional 
CP-even and/or -odd Higgs states, which can be accessible by the LHC experiments in a variety of signatures.

This paper focuses on the popular 2HDM. After EW Symmetry Breaking (EWSB), the scalar sector of the 2HDM predicts five physical Higgs states, two CP-even Higgses ($h,~H$, with $m_h<m_H$), one CP-odd one ($a$) and a pair of charged ones ($H^\pm$). The rich (pseudo)scalar sector of the 2HDM and the different sets of Yukawa couplings that can be realised then offer a very interesting production and decay phenomenology of neutral and charged Higgs states at the LHC, even after scrutinising the 2HDM parameter space by considering different theoretical (vacuum stability, perturbativity, unitarity, etc.) and  experimental (from SM-like Higgs data and nil searches for companion states, flavour physics and low energy observables, etc.) constraints. Furthermore, the 2HDM is also attractive because one can impose a simple   $Z_2$ discrete symmetry to the Yukawa sector in order to suppress  Flavour Changing Neutral Currents (FCNCs) at tree level, which then forces one doublet to couple to a given type of fermions and leading as a result to four Yukawa interactions (termed, Type-I, Type-II, Type-X and Type-Y). In fact, in order to realise EWSB in such a way that the 2HDM is compliant with all experimental data, it is finally customary to allow for a soft breaking of this $Z_2$ symmetry. Herein, we will use the latter setup with a Type-I Yukawa structure.


Specifically, in the present study, we plan to take advantage of the direct access to some trilinear Higgs couplings that the LHC can access, entering multi-Higgs processes such as $H \to hh(aa)$ and $H^\pm \to W^\pm a$, to search for light Higgs states in cascade (or chain) or decays in the framework of 2HDM Type-I. In fact, as the analysis progress, we aim to use the information and the strong correlation between the aforementioned couplings to explore the scope of a new search for light Higgses at the LHC Run 3 (with an integrated luminosity of 300 $\text{fb}^{-1}$) as well as the HL-LHC (with an integrated luminosity of 3000 $\text{fb}^{-1}$), on the basis of the knowledge acquired from the study of the aforementioned signatures. Chiefly, we focus on the case where $H$ is the observed Higgs with a mass of 125 GeV, while $h$ and $a$ are lighter, which then opens a window for non SM-like Higgs decays, such as $H \to Z^{(*)}a$. This configuration is possible in a 2HDM Type-I, in turn offering the  possibility of an alternative and new promising signal, in the form of the following cascade decays $H \to Z^{*}a \to Z^{*}Z^{*}h \to b\overline{b} \mu^- \mu^+ jj$. The main Higgs production process is via gluon fusion $gg \to H$.

In what follows, we provide a brief review of the 2HDM Type-I, in section~\ref{section2}. Then, in section~\ref{section3}, we discuss the outcome of recasting the aforementioned cascade Higgs decays in such a framework. Section~\ref{section4} is devoted to the signal-to-background analysis of the proposed new signal based on running a full Monte Carlo (MC) simulation while we finally conclude in section~\ref{section5}.

\section{2HDM Type-I}
\label{section2}
The 2HDM is one of the simplest well-motivated extensions of the SM. In
this section, we briefly review the theoretical structure of this
model. The scalar sector of the 2HDM consists of two complex $SU(2)_L$
doublets, $\Phi_1$ and $\Phi_2$, with hypercharge $Y = +1$. The most
general $SU(2)_L\times U(1)_Y$ invariant scalar potential can be
written as follows:

\begin{eqnarray}
V(\Phi_1,\Phi_2) &=& m_{11}^2 \Phi_1^\dagger\Phi_1+m_{22}^2\Phi_2^\dagger\Phi_2-[m_{12}^2\Phi_1^\dagger\Phi_2+{\rm h.c.}] \nonumber\\
&+& \frac{\lambda_1}{2}(\Phi_1^\dagger\Phi_1)^2
+\frac{\lambda_2}{2}(\Phi_2^\dagger\Phi_2)^2
+\lambda_3(\Phi_1^\dagger\Phi_1)(\Phi_2^\dagger\Phi_2)
+\lambda_4(\Phi_1^\dagger\Phi_2)(\Phi_2^\dagger\Phi_1) \nonumber\\
&+&\left\{\frac{\lambda_5}{2}(\Phi_1^\dagger\Phi_2)^2
+\big[\lambda_6(\Phi_1^\dagger\Phi_1)
+\lambda_7(\Phi_2^\dagger\Phi_2)\big]
\Phi_1^\dagger\Phi_2+{\rm h.c.}\right\}. \label{pot1}
\end{eqnarray}
Assuming  CP-conservation in the 2HDM and following the hermiticity of the scalar potential, $m_{11}^2$, $m_{22}^2$, $m_{12}^2$, $\lambda_{1,2,3,4,5,6}$  are real parameters. Invoking the described ${Z}_2$ symmetry, to avoid tree-level Higgs-mediated FCNCs at tree level, implies that $\lambda_{6} = \lambda_{7} = 0$. Also notice that the bilinear term proportional to $m_{12}^2$ breaks the $Z_2$ symmetry softly.
Using the two minimisation conditions of the scalar potential and the combination $v^2=v_1^2+v_2^2=(2\sqrt{2} G_F)^{-1}$, one can then trade the Lagrangian parameters of the 2HDM for a more convenient set of variables,
\begin{center}
	$\alpha$,\, $\tan\beta= \frac{v_2}{v_1}$,\,  $m_{h}$,\, $m_{H}$  ,\, $m_a$,\, 
	$m_{H^\pm}$\, \ \rm{and}\ \  $m_{12}^2$,
\end{center}
where $\alpha$ is the CP-even mixing angle, $v_1$ and $v_2$ are the Vaccum Expectations Values (VEVs) of the two Higgs doublets $\Phi_1$ and $\Phi_2$, respectively. 

\subsection{Yukawa couplings}
The general structure of the Yukawa Lagrangian when both Higgs fields couple to all fermions is given by:
\begin{eqnarray}
{\cal{L}}_Y &=& \bar{Q'}_L ( Y^{u}_1 {\tilde \Phi_{1}} + Y^{u}_2 {\tilde \Phi_{2}}) u'_{R}
+\bar{Q'}_L (Y^{d}_1 \Phi_{1} + Y^{d}_2 \Phi_{2}) d'_{R} 
+ \bar{L'_{L}} (Y^{l}_{1}\Phi_{1} + Y^{l}_{2}\Phi_{2}) l'_R + \rm{h.c.},
\label{yuk_III}
\end{eqnarray}
where $Q'_L$ and $L'_L$ are the weak isospin quark and lepton doublets, $u'_R$ and $d'_R$ denote the right-handed quark singlets while $Y_{1,2}^{u}$, $Y_{1,2}^{d}$ and $Y_{1,2}^{l}$ are couplings matrices in flavour space. 
This form of Yukawa interaction gives rise to large FCNCs at tree level, which is strongly constrained by $B$-physics observables. Implementing $Z_2$ symmetry allows only one doublet to couple to a given right-handed fermion field. Depending on the $Z_2$ assignment, one can have the four types of models previously refered to as  Type-I, Type-II, Type-X and Type-Y. In the mass-eigenstate basis, they can be unified and expressed as follows:
\begin{align}
-{\mathcal L}_Y=&+\sum_{f=u,d,\ell} \left[m_f \bar f f+\left(\frac{m_f}{v}\kappa_h^f \bar f fh+\frac{m_f}{v}\kappa_H^f \bar f fH-i\frac{m_f}{v}\kappa_A^f \bar f \gamma_5fA \right) \right]\nonumber\\
&+\frac{\sqrt 2}{v}\bar u \left (m_u V \kappa_A^u P_L+ V m_d\kappa_A^d P_R \right )d H^+ +\frac{\sqrt2m_\ell\kappa_A^\ell}{v}\bar\nu_L \ell_R H^+
+\text{h.c.},\label{Eq:Yukawa}
\end{align}
where $P_{L,R}=(1\pm \gamma_5)/2$ are the projection operators and $V$ denotes the Cabibbo-Kobayashi-Maskawa (CKM) matrix. 

Here, we focus only on Type-I, where only one doublet $\Phi_2$ couples to all fermions, and thus the Higgs-fermion couplings are flavour diagonal in the fermion mass basis and depend only on the mixing angles, $\alpha$ and $\beta$, as follows: 
\begin{eqnarray}
\kappa^{u,~d,~l}_{h} &=&c_\alpha/ s_\beta ~=~ s_{\beta-\alpha}+\cot \beta~c_{\beta-\alpha},   \\ 
\kappa^{u,~d,~l}_{H} &=&s_\alpha/ s_\beta~=~c_{\beta-\alpha}-\cot \beta~s_{\beta-\alpha}, \\
\kappa^{d,~l}_{A}&=& -\cot \beta,~~ \kappa^{u}_{A}= \cot \beta,   
\end{eqnarray}
where we have used the short-hand notation $c$ and $s$ for $\cos$ and $\sin$, respectively.

\subsection{Theoretical and Experimental Constraints}
\label{constraint}
We now describe briefly a set of, in turn,  theoretical and experimental constraints that must be satisfied by the parameter space of the 2HDM.
\begin{itemize}
\item[\textbullet]Perturbative unitarity~\cite{Kanemura:1993hm, Akeroyd:2000wc,Arhrib:2000is} forces several constraints on the quartic couplings of the scalar potential by requiring the following inequalities:\\
\begin{eqnarray}
\frac{3}{2} \left\{
(\lambda_1 + \lambda_2) \pm
\sqrt{ (\lambda_1-\lambda_2)^2 + \frac{4}{9} (2\lambda_3+\lambda_4)^2}
\right\}< ~8\pi, \ ~ \
\left(
\lambda_3 \pm \lambda_5 \right)< ~8\pi,  
\\ 
\frac{1}{2} \left\{
(\lambda_1 + \lambda_2) \pm
\sqrt{ (\lambda_1-\lambda_2)^2 +4 \lambda_4^2} 
\right\}<  ~8\pi, \ ~ \ \left(
\lambda_3 + 2 \lambda_4 \pm 3 \lambda_5 \right)< ~8\pi,  \\
\frac{1}{2} \left\{
(\lambda_1 + \lambda_2) \pm
\sqrt{(\lambda_1-\lambda_2)^2 +4 \lambda_5^2}
\right\}< ~8\pi.
\end{eqnarray}

\item[\textbullet]Vacuum stability~\cite{ElKaffas:2006gdt} requires the scalar potential to be finite at large field values and this can be translated into these bounds:
\begin{eqnarray}
\lambda_{1,2}>0,  \,\,
\lambda_3>- \sqrt{\lambda_1\lambda_2}, \,\,
\lambda_3+\lambda_4-|\lambda_5|> - \sqrt{\lambda_1\lambda_2}.
\label{eq:VB}
\end{eqnarray}	

\item[\textbullet]Perturbativity  requires the quartic couplings to obey $|\lambda_i| < 4\pi$ ($i=1,\ldots,5$). 		
\end{itemize}
Experimental observations impose the following constraints:

\begin{itemize}
\item[\textbullet]EW precision observables, i.e.,  the oblique parameters, $S, T$ and $U$~\cite{Peskin:1990zt, Peskin:1991sw} are required to be  within 95\% C.L. of their experimental measurements, the current fit values (with the correlation parameters) are given by~\cite{particle2020review}:
\begin{eqnarray}
S= -0.01\pm 0.10,\  \ T = 0.03\pm 0.12,\ \ U = 0.02\pm 0.11,  \nonumber \\
\rho_{ST} = 0.92,\ \ \rho_{SU} = -0.80,\ \ \rho_{TU} = -0.93,~~ \chi_{ST,~SU,~TU} < 5.99. \nonumber 
\end{eqnarray}
\end{itemize}

The above constraints have been implemented in \texttt{2HDMC-1.8.0}~\cite{2hdmc}. This public code is then used to scan over the parameter space of the 2HDM and test it against the above contraints as well as to compute the different Higgs BRs in each point of it. (\texttt{2HDMC} also provides an interface to \texttt{HiggsBounds} and \texttt{HiggsSignals}, see below.)

Further experimental observations are utlized as follows:


\begin{itemize}
	\item[\textbullet]Consistency with the $Z$ width measurement $\Gamma_Z = 2.4952 \pm 0.0023$ GeV from LEP~\cite{particle2020review} is required. 

	\item[\textbullet] Constraints from LHC, Tevatron and LEP searches which failed to find companion Higgs states are taken into account via  \texttt{HiggsBounds-5.10.0}~\cite{Bechtle:2020pkv}, which allows to test the exclusion limits at 95\% C.L.  

\item[\textbullet] The code 
\texttt{HiggsSignals-2.6.2}~\cite{Bechtle:2020uwn} is used to check the signal strength measurements of the SM-like Higgs boson discovered at the LHC in 2012. 

	\item[\textbullet] Constraints from $B$-physics observables are enforced by Superiso-v1.4~\cite{Mahmoudi:2008tp}, using the following measured observables:
	\begin{eqnarray}
	{\rm BR}(B_d \to \mu^+ \mu^-) = (3.9\pm 1.5)\times 10^{-4}~\text{\cite{Haller:2018nnx}},& \  & {\rm BR}(B_s \to \mu^+ \mu^-) = (3.0\pm 0.6)\times 10^{-9}~\text{\cite{LHCb:2017rmj}}, \nonumber \\ 
	{\rm BR}(B \to X_s \gamma)& =& (3.32\pm 0.15)\times 10^{-3}~\text{\cite{HFLAV:2016hnz}}.  \nonumber 
	\end{eqnarray}
	
	\item[\textbullet]Constraints from recent searches for light pseudoscalar states in the mass range $[15,~62.5]$ GeV, in proton-proton collision at $\sqrt{s}=13~$ TeV, in $2\mu2b$~\cite{CMS:2018nsh,ATLAS:2018emt}, $2\tau 2\mu$~\cite{CMS:2018qvj} and $2b2\tau$\cite{CMS:2018zvv} final states, are included in \texttt{HiggsBounds}. Since no significant excess is observed, upper limits are set on ${\rm BR}(H \rightarrow  aa \rightarrow 2\mu 2b,~2\mu 2\tau,~2b2\tau )$~\cite{CMS:2018nsh,CMS:2018qvj,CMS:2018zvv}. However, lately, additional constraints from such Higgs cascade decays have emerged, not included in the numerical tool, so we had to deal with these separately. For example,  
the CMS group has reported a search for $H \to aa \to 4 \gamma$~\cite{CMS:2021bvh}, using the data collected at $\sqrt{s} =13$ TeV, with an integrated luminosity of 132 fb$^{-1}$. Upper limits can then be set on ${\rm BR}(H \to aa \to 4\gamma)$ at 95\% C.L, since no significant deviation is observed\footnote{EasyNData~\cite{Uwer:2007rs} was used to digitise the exclusion limits from the published papers in order to test each point in the parameter space against the upper limit on ${\rm BR}(H \to hh(aa) \to 4\gamma)$~\cite{CMS:2021bvh}, a procedure which was validated against the case of ${\rm BR}(H \to hh(aa) \to 2\mu 2\tau)$ using \cite{CMS:2020ffa}.}.
    The ATLAS group~\cite{ATLAS:2021hbr} has also recently searched for the exotic decay of the Higgs boson into two light pseudoscalars in $2b2\mu$ final state at $\sqrt{s}=13$ TeV with an integrated luminosity of 137 fb$^{-1}$, in the range of masses varying from 15 GeV to 60 GeV. (The largest excess with a local significance  of $3.3\sigma$ is observed at a dimuon invariant mass of 52 GeV.) In the background only hypothesis, upper limits at 95\% C.L. can  be placed on ${\rm BR}(H \to aa \to 2b2\mu)$\footnote[3]{Corresponding search data and exclusion limits are available at the HEPData database.}.  Tab.~\ref{Tab2} summarises several searches for exotic decays of the Higgs bosons in various final states, performed by the two collaborations ATLAS and CMS at Run 2, targeting a different ranges of masses. 
\begin{table}[h!]
	\begin{center}
			\resizebox{0.75\textwidth}{!}{
			\begin{tabular}{|c |c|c|c|} \hline\hline
				Limit & Collaboration & Range & HiggsBounds\\ 
				\hline \hline
				$S \to HH \to 2b2\gamma$~\cite{CMS:2018tla}& CMS & $250\text{GeV}< m_S < 900$  & $\times$ \\
				\hline
				$S \to HH \to 2b2\tau$~\cite{CMS:2017hea}& CMS & $~250\text{GeV}< m_S < 900 \text{GeV}~$ & $\times$ \\
				\hline
				$H \to aa \to 2b2\mu$~\cite{CMS:2018nsh}& CMS & $15\text{GeV}< m_a < 60 \text{GeV}$ & \checkmark \\
					\hline
				$H \to aa \to 2b2\mu$~\cite{ATLAS:2018emt} & ATLAS & $15\text{GeV}< m_a < 60 \text{GeV}$ & \checkmark \\
				\hline
				$H \to aa \to 2\mu 2\tau$~\cite{CMS:2018qvj} & CMS & $15\text{GeV}< m_a < 61.5 \text{GeV}$ & \checkmark \\
				\hline
				$H \to aa \to 2b2\tau$~\cite{CMS:2018zvv}& CMS & $15\text{GeV}< m_a < 60 \text{GeV}$ & \checkmark \\
								\hline
				$H \to aa \to 4\gamma$~\cite{CMS:2021bvh}& CMS & $15\text{GeV}< m_a < 60 \text{GeV}$ & \textcolor{black}{$\times$} \\
					\hline
			     $H \to aa \to 2\mu 2\tau$~\cite{CMS:2020ffa} & CMS & $3.6\text{GeV}< m_a < 21 \text{GeV}$ & \textcolor{black}{$\times$}\\
				\hline
				$H \to aa \to 2b2\mu$~\cite{ATLAS:2021hbr} & ATLAS & $15\text{GeV}< m_a < 60 \text{GeV}$ & \textcolor{black}{$\times$}    \\
				\hline
				$~S \to HH\to bbVV$~\cite{CMS:2017rpp},~\cite{CMS:2019noi} & CMS & $~260\text{GeV}< m_S < 900 \text{GeV}~$ & \checkmark \\
				\hline
				$H \to aa \to 4b$~\cite{ATLAS:2018pvw}& ATLAS & $20\text{GeV}< m_a < 60 \text{GeV}$ & \checkmark \\
				\hline
				$S \to HH \to 2b2\gamma$~\cite{ATLAS:2018dpp}&  ATLAS & $260\text{GeV}< m_S < 1000$  & \checkmark \\
				\hline
				$S \to HH \to 2b2\gamma$~\cite{ATLAS:2021jki}&   ATLAS&$250\text{GeV}< m_S < 1000$ &$\times$\\
				\hline				
		\end{tabular} 
	             }          
	\end{center}
\caption{$(\times)$/(\checkmark) indicate searches (not yet)/(already) implemented in \texttt{HiggsBounds-5.10.0}.}
   \label{Tab2}
\end{table}	 
\end{itemize}	

\section{Numerical Analyses}
\label{section3}
The (pseudo)scalar sector of the 2HDM involves two CP-even Higgses, $h$ and $H$. One of these scalars can be identified as the  125 GeV state observed at the LHC. As mentioned, in this analysis, we will assume that the heaviest Higgs state $H$ is the SM-like one with a mass of 125 GeV and that $h$ and $a$ are lighter than $H$. We the perform a scan over the following ranges,
\begin{eqnarray*}
 m_h \in [10~\text{GeV},~90~\text{GeV}],~~~m_H = 125~\text{GeV},~~~m_a \in [10~\text{GeV},~90~\text{GeV}], \nonumber \\
 m_{H^\pm}\in [100~\text{GeV},~160~\text{GeV}],~~~\tan\beta \in  [2.5,~25],~~~\sin(\beta-\alpha) \in [-0.7,~0.0],
 \label{eq1}
\end{eqnarray*}
with $m_{12}^2$= $m_a^2\tan\beta/(1+\tan^2\beta)$. Assuming  $m_H = 125$ GeV and $m_{h,a} < 90$ GeV, the decay channels $H \to h h,aa,~aZ^{*}$ could be open, leading to invisible or undetected SM-like Higgs decays that are restricted by the current precision measurements of Higgs couplings. CMS performed a combination of searches, using data collected at $\sqrt{s} = 7,~8,~13$ TeV~\cite{CMS:2018yfx}, for Higgs bosons decaying into invisible particles, which targets the following production channels:  Vector Boson Fusion (VBF), Higgs-Strahlung (HS) and gluon-gluon Fusion (ggF) (allowing for  initial state radiation). The combination  of all the searches, assuming these SM-like production modes, yields an observed (expected) upper limit on ${\rm BR}(H \to inv)$ of 0.19 (0.15) at 95\% C.L. The ATLAS group reported a direct search for Higgs bosons produced via VBF with subsequent invisible decays, for 139 $\text{fb}^{-1}$ of $pp$ collision data at $\sqrt{s} = 13$ TeV~\cite{ATLAS:2022yvh}. An observed (expected) upper limit of 0.145 (0.103) is placed on ${\rm BR}(H \to inv)$ at 95\% C.L., as a function of the assumed production cross sections. In our analysis, we will assume that ${\rm BR}(H \to inv)$ designates the sum of the following decay rates, ${\rm BR}(H \to hh)$, ${\rm BR}(H\to aa)$ and ${\rm BR}(H \to aZ^{*})$. 

\begin{figure}[!h]
	\centering
	\resizebox{0.7\textwidth}{0.3\textheight}{
		\includegraphics{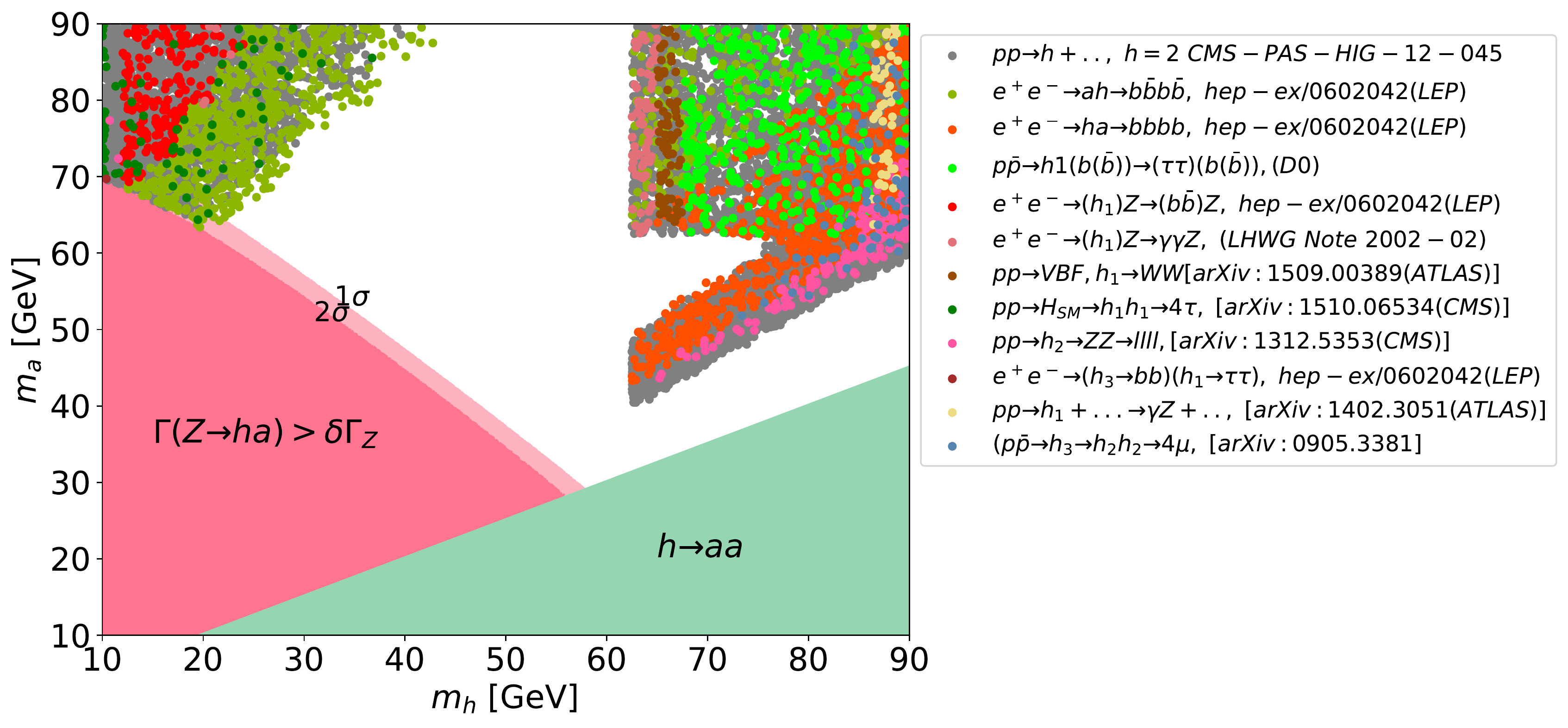}}
	\caption{Allowed parameter space in the 2HDM Type-I  at 95\% C.L. Coloured dots represent the searches to which the relevant ($m_h, m_a$) regions are sensitive to.}
	\label{fig1}
\end{figure}

After performing a random scan over 2HDM Type-I parameters, we show in
Fig.~\ref{fig1} the region allowed by theoretical and experimental
constraints. Within this region the most sensitive channels to the parameter space
of the model as determined by \texttt{HiggsBounds} are shown by coloured dots.
Obviously, there are two distinct regions in the figure. The one in the top left
corner corresponds to low masses of $h$ ($m_h < m_H/2$), and high
masses of $a$ ($m_a> m_H/2$), while the second one corresponds to the
$m_{a,~h}> m_H/2$ scenario. It is interesting to note that there are
no acceptable points when $40~\text{GeV}<m_h<m_H/2$ and $m_a >
m_H/2$. This is due to the fact that this parameter combination is
excluded by LEP searches for $e^+e^- \to ah \to
b\overline{b}b\overline{b}$~\cite{ALEPH:2006tnd} and an ATLAS search
for events with at least $3\gamma$ in $pp \to H_{\rm SM} \to h h \to 4\gamma$~\cite{ATLAS:2015rsn}.
The figure also captures the
constraints from the $Z$ width, which forbid possible mass
combinations $(m_h,~m_a)$ when $\cos(\beta-\alpha)\to
1$. Additionally, the constraint from the LEP search for the $e^+ e^-
\to (h \to aa)a \to (b\overline{b}b\overline{b})b\overline{b}$
process~\cite{ALEPH:2006tnd} excludes the bottom right region
corresponding to $m_h > 2m_a$, where the decay channel $h \to aa$ is
kinematically open.

Finally, 
an interesting observation is that the sensitivity in the region with low $m_h$ and high $m_a$ is mainly from LEP searches for processes such as $e^+e^- \to ah \to b\overline{b}b\overline{b}$ and $e^+ e^- \to (h)Z \to (b\overline{b})Z$~\cite{ALEPH:2006tnd}. Therefore, an update from the LHC during Run 3 is unlikely to rule out this mass combination over the plane $(m_h,~m_a)$ of the 2HDM Type-I. We will be focusing on this region in the second part of our study. 

\begin{figure}[h!]
	\resizebox{0.5\textwidth}{!}{
		\includegraphics{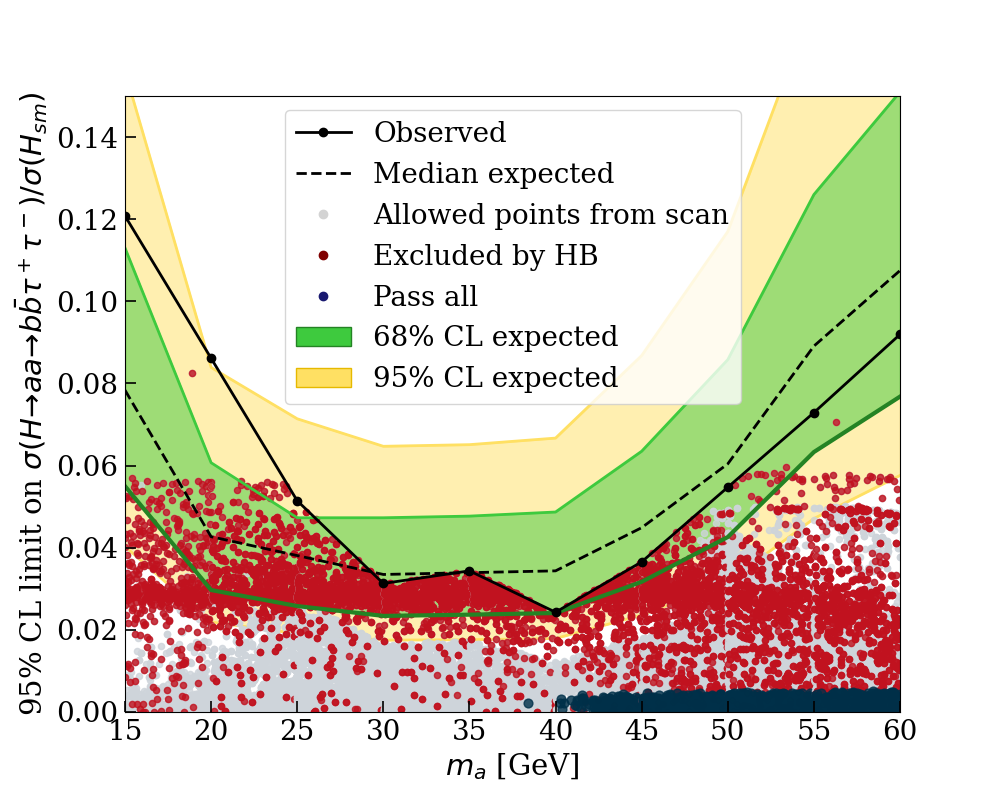}}
	\resizebox{0.5\textwidth}{!}{
		\includegraphics{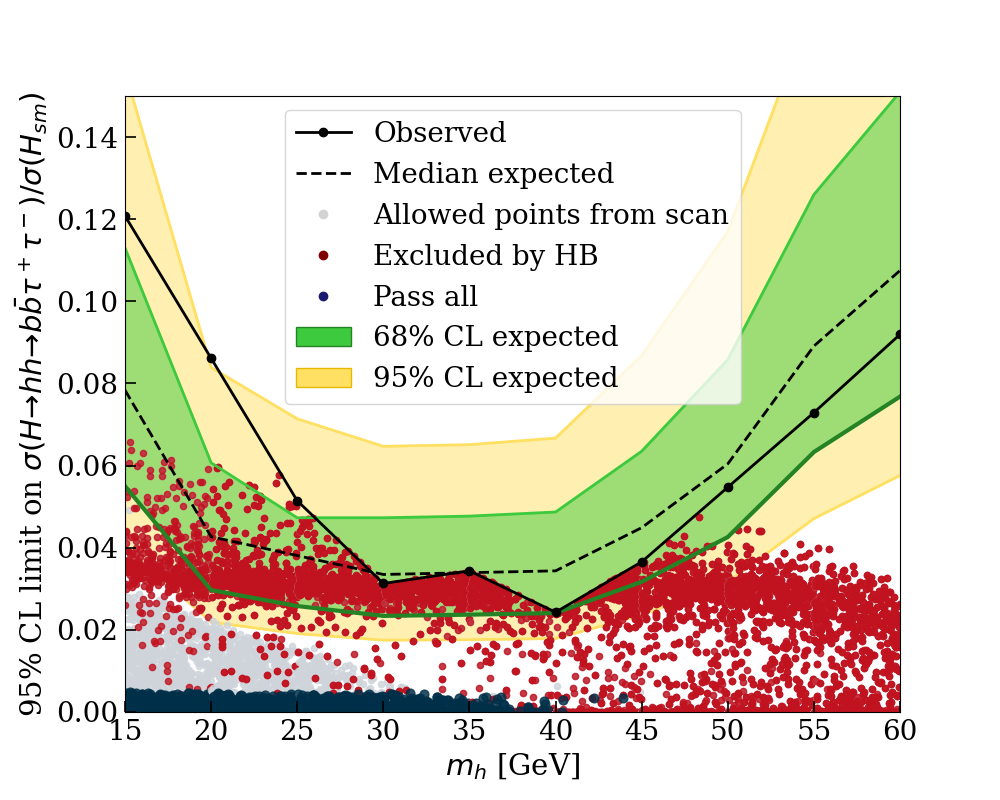}}
	\caption{Observed and expected upper limits on $\sigma(H \rightarrow aa (hh) \rightarrow b\overline{b} \tau^+\tau^-)/\sigma_{\rm SM}(H)$~\cite{CMS:2018zvv} at 95\% C.L.  in the 2HDM Type-I. Grey points are allowed by theoretical constraints. Blue points satisfy both theoretical and experimental constraints} 
	\label{fig2}
\end{figure}

We now turn to the reinterpretation of exotic Higgs decay searches, i.e., $H \to aa$ in $2b2\tau$, $2b2\mu$ and $2\mu2\tau$ final states in the framework of the 2HDM Type-I, while taking advantage of the parameter space discussed above.
The recasting of  $2b2\tau$, $2b2\mu$ and $2\tau2\mu$ searches for $H\to hh$ is also possible since these processes share similar kinematics (in the same spirit as in Ref.~\cite{Semlali:2020cxl}). It is relevant to note that the constraints from the search for light pseudoscalars in the $2b2\tau$ final state are much stronger than the ones from $ 2b2\mu$ and $2\tau2\mu$ searches. CMS has set an upper limit, between 3\% and 12\%, on ${\rm BR}(H \to aa \to 2b2\tau)$  at 95\% C.L.~\cite{CMS:2018zvv}, assuming the SM production of primary Higgs boson. We show in Fig.~\ref{fig2} the outcome from reinterpreting the $H \to aa (hh) \to 2b2\tau$ search in the 2HDM Type-I. Grey points satisfy theoretical constraints as described in section~\ref{constraint}, whereas red points are excluded by nil searches (i.e., by \texttt{HiggsBounds}). The area of sensitivity to $H \to aa (hh) \to 2b2\tau$ is excluded. The blue points satisfy both theoretical and experimental constraints. In this connection,  the BR of Higgs SM-like Higgs state decaying into $hh$ and/or $aa$ is very restricted and cannot exceed $9\%$ at 95\% C.L., again, in the 2HDM Type-I. 
\begin{figure}[h!]
	\resizebox{0.5\textwidth}{!}{
		\includegraphics{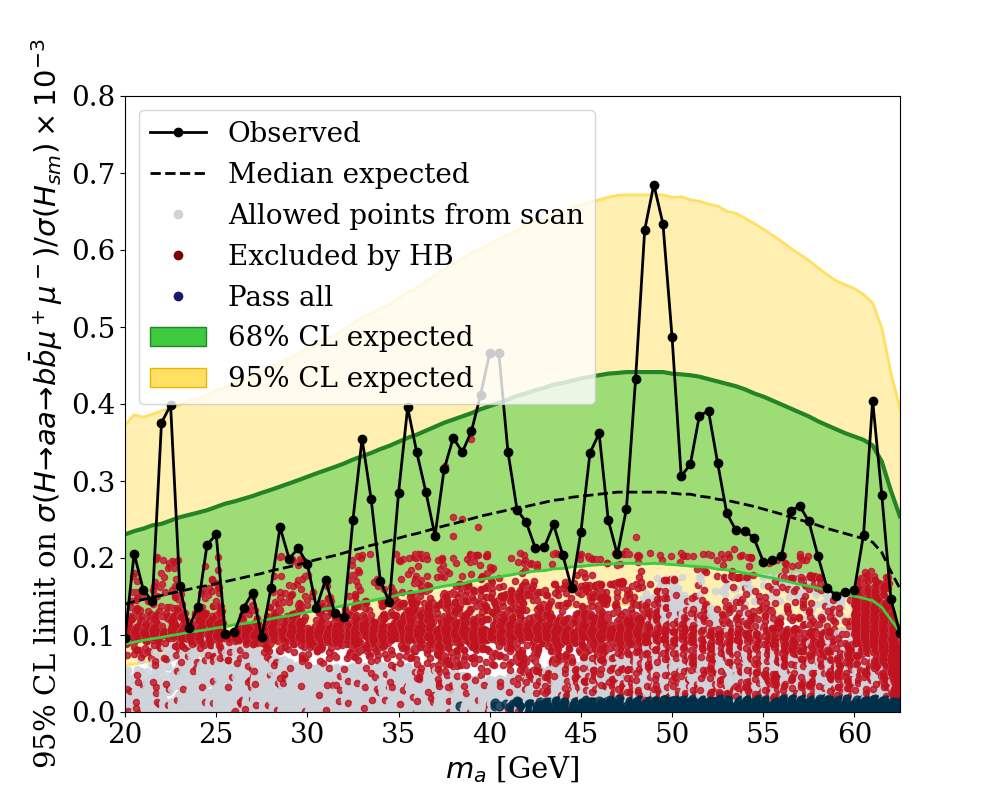}}
	\resizebox{0.5\textwidth}{!}{
		\includegraphics{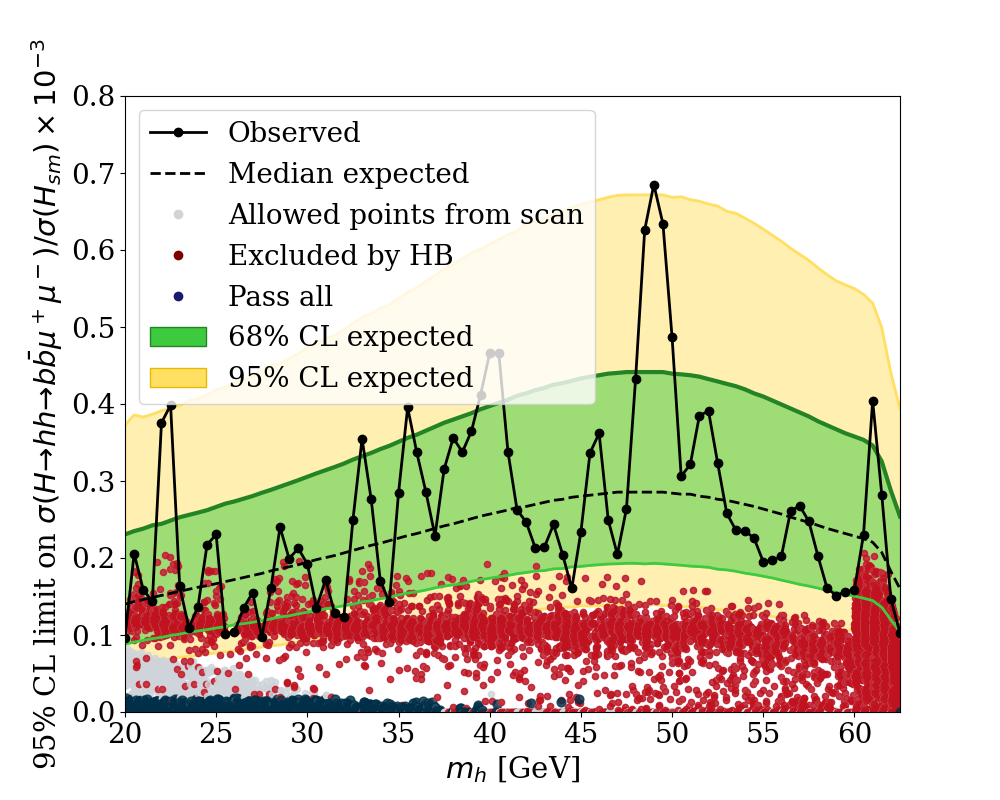}}
	\caption{Observed and expected upper limits on $\sigma(H \rightarrow aa (hh) \rightarrow b\overline{b} \mu^+ \mu^-)/\sigma_{\rm SM}(H)$~\cite{CMS:2018nsh} at 95\% C.L.  in the 2HDM Type-I.} 
	\label{figg1}
\end{figure}
%
One can draw a similar conclusion form reinterpreting $H \to hh(aa) \to 2b2\mu$  in our reference framework. Fig.~\ref{figg1} shows that the parameter space with sensitivity to this search is excluded . One should keep in mind that the $2b2\mu$ final state is well-balanced between large ${\rm BR}(h/a \to b\overline{b})$ and a clean di-muon resonance that is easy to trigger on. This exercise emphasises that the 2HDM Type-I may not be a good framework for reinterpreting searches for exotic Higgs decays into light pseudoscalar in   ``traditional''  final states such as $2b2\mu$, $2b2\tau$ and $2\tau2\mu$.
We also address here light charged Higgs decay in the mass ranges where $m_{H^\pm} < m_t-m_b$ and $m_{h,a} < 90~\text{GeV}$. In this configuration, the charged Higgs state can be be produced from top quark decays, i.e., $t \to b H^+$, followed by its bosonic decays to $H^\pm \to W^\pm h (a)$, instead of the standard fermionic decay  modes like $\tau \nu$ and $cs$. Many studies  motivated these channels as alternative modes to search for light charged Higgs bosons that could dominate over the conventional fermionic channels, because of large BRs when they are kinematically allowed,  in models such as our 2HDM Type-I~\cite{Arhrib:2016wpw,Arhrib:2021xmc,Wang:2021pxc}. ATLAS~\cite{ATLAS:2021xhq} and CMS~\cite{CMS:2019idx} have considered the ranges $m_a \in[15,75]$ GeV and $m_{H^\pm} < m_t-m_b$ to search for light charged Higgs bosons in $pp \to t \overline{t} \to b\overline{b}H^+ W^-$ with $H^+ \to W^+ a$ and $a \to \mu^+ \mu^-$ at $\sqrt{s}=13~\text{TeV}$, since the $\mu^+\mu^-$ finale state provides the aforementioned experimental advantages, which offset the suppressed rate of ${\rm BR}(a \to \mu^+ \mu^-)$. Previously, both CDF and the LEP collaborations have searched for $H^\pm \to W^\pm a$ with $a \to b\overline{b}$~\cite{CDF:2005acr}, $a \to \tau^+\tau^-$~\cite{CDF:2011pxh} and $a \to b\overline{b}$~\cite{OPAL:2008tdn}. In addition, LEP experiments~\cite{ALEPH:2013htx} have set a lower bound on the charged Higgs boson mass of $m_{H^\pm} > 72.5~\text{GeV}$ in the 2HDM Type-I for $m_a>12~\text{GeV}$ at 95\% C.L.
\begin{figure}[h!]
	\centering
	\resizebox{0.5\textwidth}{!}{
		\includegraphics{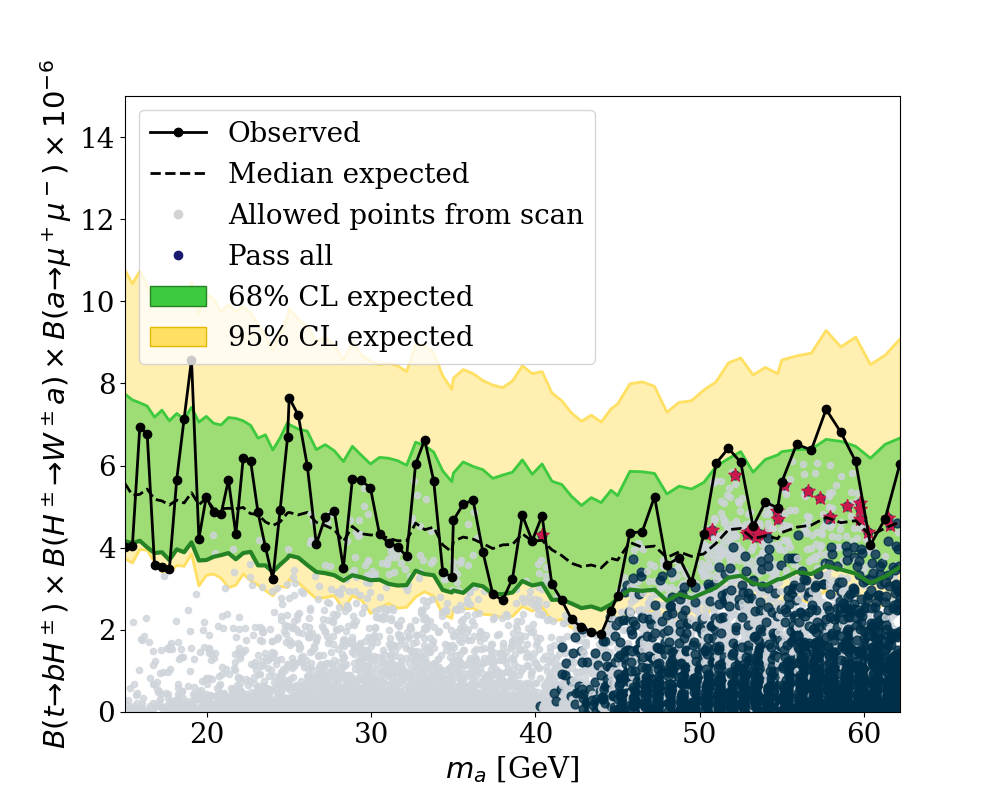}}
	\caption{Observed and expected upper limits on ${\rm BR}(t \to H^+b) \times  {\rm BR}(H^+ \to W^+ a) \times {\rm BR}(a \to \mu^+ \mu^-)$~\cite{CMS:2019idx} at 95\% C.L.  in the  2HDM Type-I.} 
	\label{fig3}
\end{figure}

Fig.~\ref{fig3} shows the CMS observed and expected exclusion limits on the product of the BRs of $t \to bH^\pm,~H^\pm \to W^\pm a$ and $a \to \mu^+\mu^-$~\cite{CMS:2019idx} as a function of $m_a$ predicted by the 2HDM Type-I, with respect to several theoretical and experimental constraints. We adopt here $m_{H^\pm} = m_a + 85$ GeV~\cite{CMS:2019idx}, which enables us to consider  $H^\pm \to W^{\pm(*)} a$, with $W^{\pm(*)}$ being on/off shell, by randomly sampling values of the charged Higgs mass between 100 GeV and 160 GeV (see Eq.~(\ref{eq1})). The yellow and green bands represent the uncertainties at $\pm 1\sigma$ and $\pm 2\sigma$ associated with the expected exclusion limits. An interesting observation is that the 2HDM Type-I offers sufficient sensitivity, when the prediction of the model exceeds the expected limit produced at $\sqrt{s} = 13$ TeV with an integrated luminosity of $35.9~\text{fb}^{-1}$ (purple stars). Such a signature could be exploited to search for a light $H^\pm$ at future experiments, Run 3 and/or the HL-LHC, given the available energies and luminosities by then. Therefore, we present in Tab.~\ref{Bp} some Benchmark Points (BPs) to test the actual sensitivity of these experiments to the 2HDM Type-I parameter space.    
\begin{table}[h!]
	\begin{center}
		\setlength{\tabcolsep}{30pt}
		\renewcommand{\arraystretch}{0.8}
		\begin{adjustbox}{max width=\textwidth}		
			\begin{tabular}{lcccc}
				\hline\hline
				Parameters &       BP$1$ &       BP$2$ &       BP$3$ &       BP$4$  \\\hline\hline
				\multicolumn{5}{c}{(Masses are in GeV)} \\\hline
				$m_h$   &    62.86&    75.69 &    75.58 &    77.18 \\
				$m_H$   &   125 &   125 &   125 &   125 \\
				$m_a$   &    40.37 &   50.73&    52.90 &    53.44 \\
				$m_{H^\pm}$   &   105.19 &   108.15 &   110.83 &   111.95\\
				$\tan\beta$ &    4.82 &     4.73 &     4.58 &     4.57 \\
				$\sin(\beta-\alpha)$  &    $-0.203$ &    $-0.209$ &    $-0.220$ &   $ -0.0.215$ \\
				\hline\hline\multicolumn{5}{c}{Total decay width in GeV} \\\hline 
				$\Gamma(h)$   &  $1.9\times10^{-6}$ &  $3.00\times10^{-6}$ &  $1.9\times10^{-6}$ &  $3.00\times10^{-6}$ \\
				$\Gamma(H)$  &  $4.54\times 10^{-3}$ &  $4.53\times 10^{-3}$ &  $4.47\times 10^{-3}$ &  $4.48\times 10^{-3}$ \\
				$\Gamma(A)$  & $5.39\times 10^{-5}$ &  $6.79\times 10^{-5}$ &  $7.6\times 10^{-5}$ &  $7.7\times 10^{-5}$\\
				$\Gamma(H^{\pm})$ &  $3.31\times 10^{-4}$ &  $3.330\times 10^{-4}$ &  $3.339\times 10^{-4}$ &  $3.339\times 10^{-4}$\\
				\hline\hline	\multicolumn{5}{c}{${\rm BR}(A\to XY)$} \\\hline
				${\rm BR}(A\to \mu\mu)$   &   $2.36\times10^{-4}$ &   $2.42\times10^{-4}$ &  $2.43\times10^{-4}$ &   $2.43\times10^{-4}$ \\				
				\hline\hline\multicolumn{5}{c}{${\rm BR}(H^{\pm}\to XY)$ in \%} \\\hline
				${\rm BR}(H^{\pm}\to W^+A)$   & 86.65  &  90.64 & 88.47  & 89.39  \\
	                 \hline\hline				
			\end{tabular}
		\end{adjustbox}
	\end{center}
	\caption{BPs  in the 2HDM Type-I.}\label{Bp}
\end{table}
In particular, we
 move now to discuss a new analysis, where we deploy the parameter space of the 2HDM Type-I following the outcomes of reinterpreting previous searches for light Higgses, $pp\to H_{\rm SM}\to hh(aa)$, in different final states, in order to search for a new signature. 	
\begin{figure}[h!]
	\centering
	\includegraphics[width=0.48\textwidth,height=0.36\textwidth]{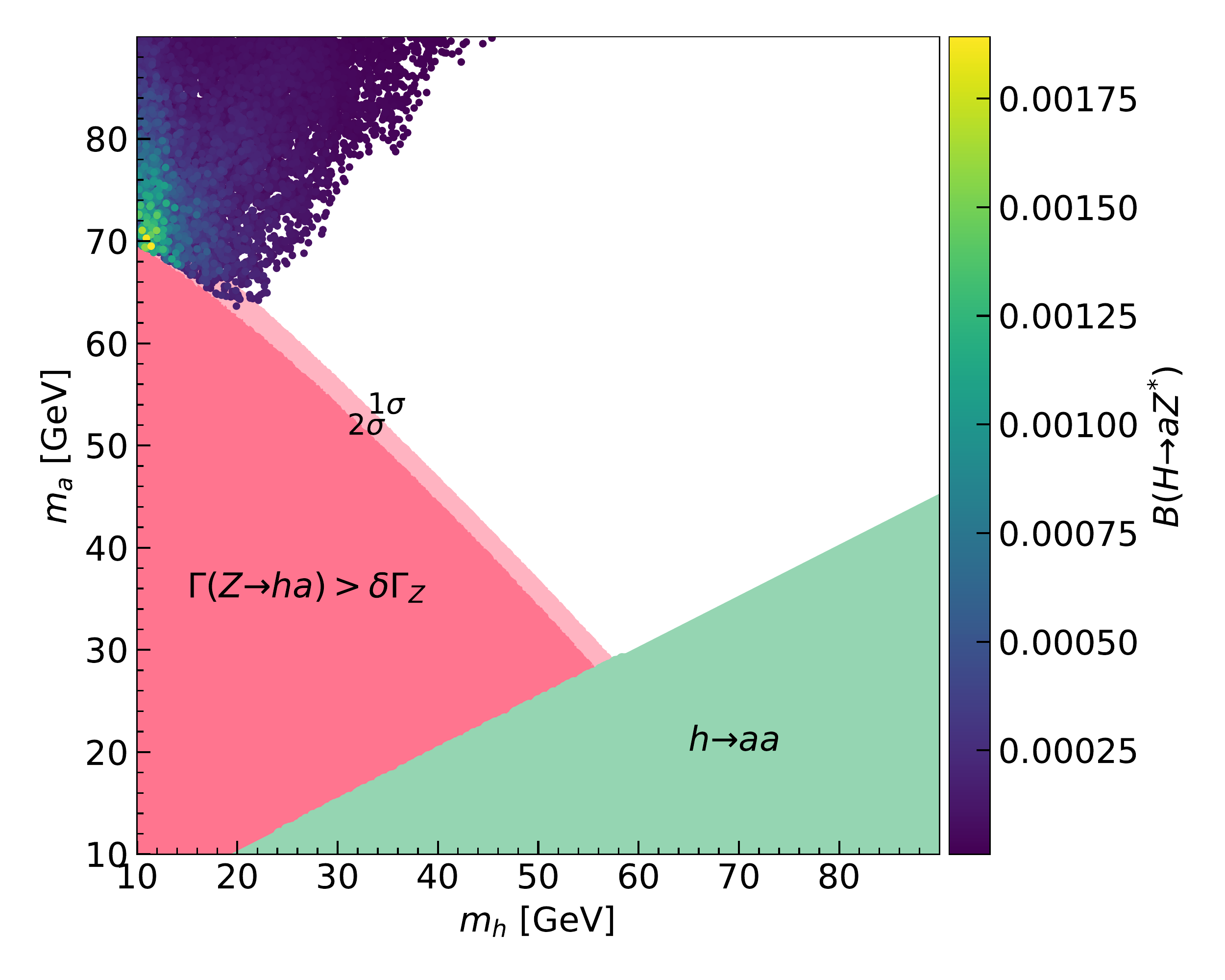}
	\includegraphics[width=0.48\textwidth,height=0.36\textwidth]{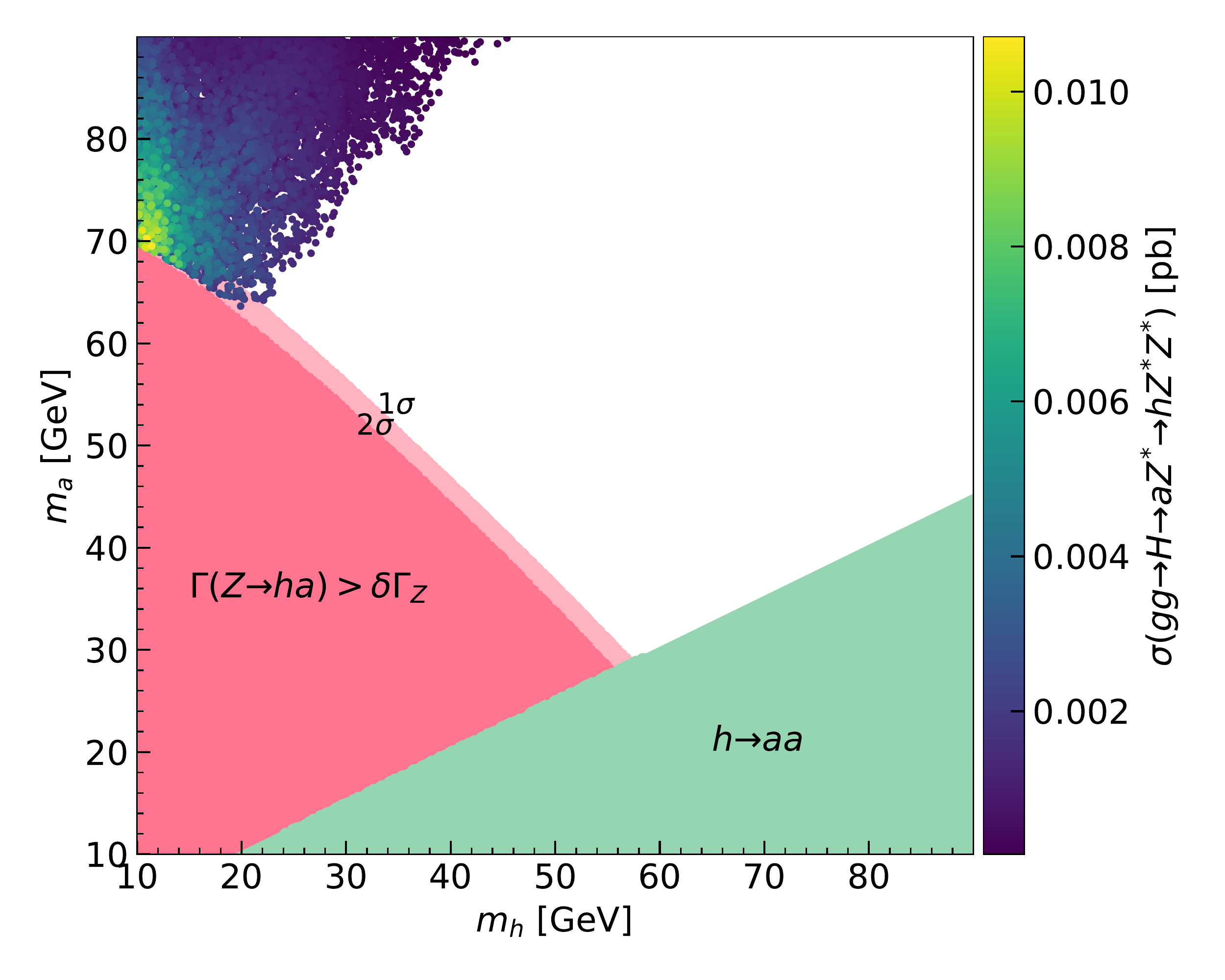}	
	\caption{$m_h$ and $m_a$ vs. ${\rm BR}(H \to Z^{*}a)$ (left) and $\sigma(gg \to H \to Z^{*}a \to Z^{*}Z^{*}h)$ (right) at 95\% C.L. in the 2HDM Type-I.}
	\label{fig5}
\end{figure}
	
Fig.~\ref{fig5} shows the result of performing a scan over the parameter space of 2HDM Type I, wherein (recall) the heaviest Higgs state is identified as the discovered SM-like one. Each sampled point is required to satisfy the theoretical and experimental constraints described in section~\ref{constraint}. In the left panel, we illustrate $m_a$ vs. $m_h$  with the BR of $H\to a Z^{(*)}$ on the colour gauge. Since $ m_H/2 < m_a < 125~\text{GeV}$, $H \to a Z^{(*)}$ will proceed with  $Z$ being off-shell, which explains the suppressed BR ($<0.2\%$). In this configuration, $H \to aa$ will not be open, thus, $H \to hh$ would only contribute significantly to the undetected decays of $H$. It should be pointed out that the total amount of ${\rm BR}( H \to aa^{*}+ aZ^{*}+ hh)$ should not exceed 19\% as required by ${\rm BR}(H \to inv)$. In the right panel, we show $m_a$ as a function of $m_h$  with $\sigma(H \to aZ^{*} \to h Z^{*}Z^{*} \to Z^{*}Z^{*})$ on the colour gauge. Once the decay chain $H \to a Z^{*}$ is open, the subsequent decay of $a$ could lead to $a \to Z^{*}h$ with $Z$ being off-shell and $h$ decaying to fermions and/or $\gamma \gamma$. We use \texttt{Sushi}~\cite {Harlander:2012pb,Harlander:2016hcx,Harlander:2002wh} to compute the cross section of Higgs production at LO\footnote{The signal cross sections is computed at LO (i.e, tree level) here, however, we will consider  QCD corrections through $K$-factors later on  in our study.}.

\begin{figure}[h!]
	\centering
	\includegraphics[width=0.48\textwidth,height=0.36\textwidth]{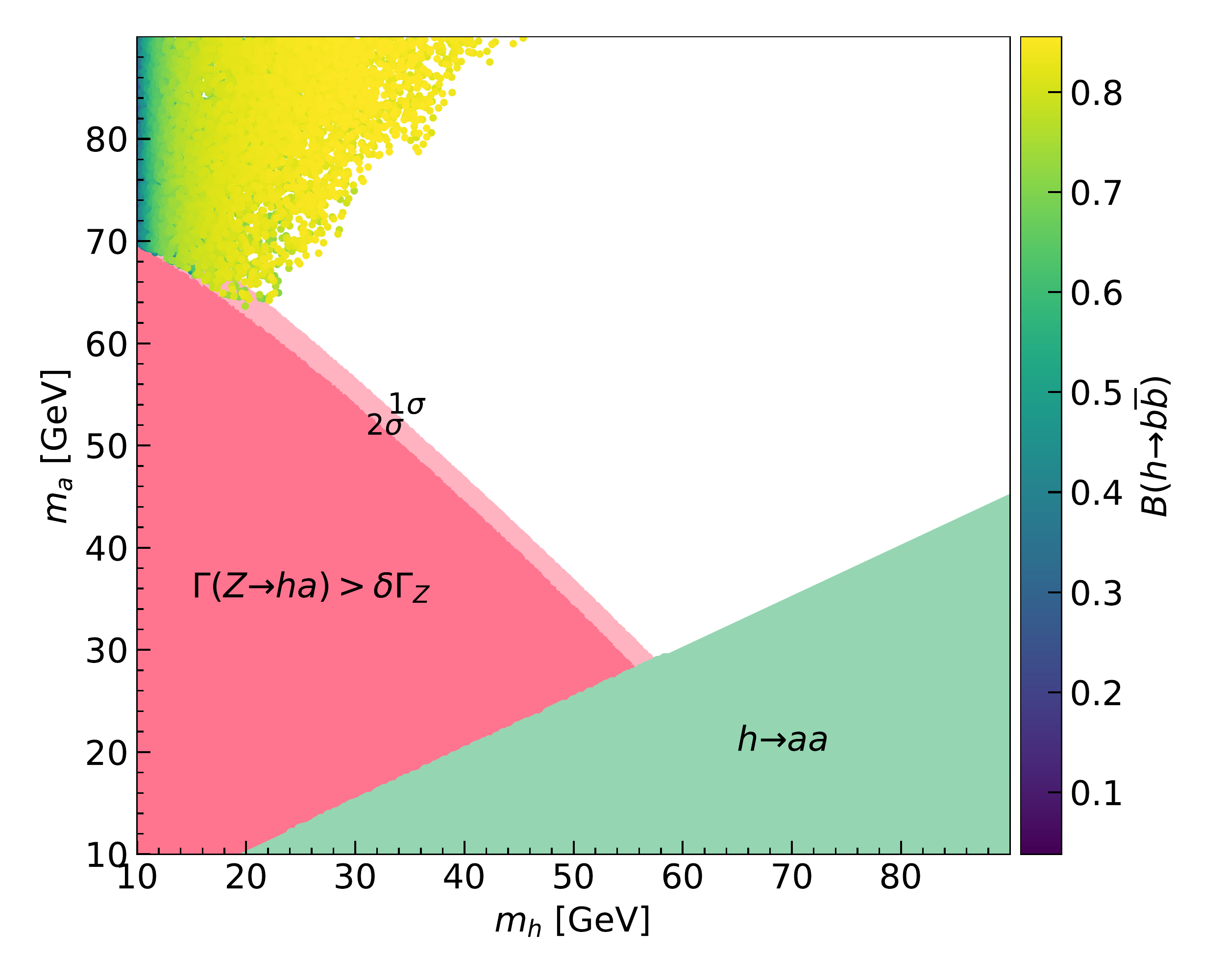}
	\includegraphics[width=0.48\textwidth,height=0.36\textwidth]{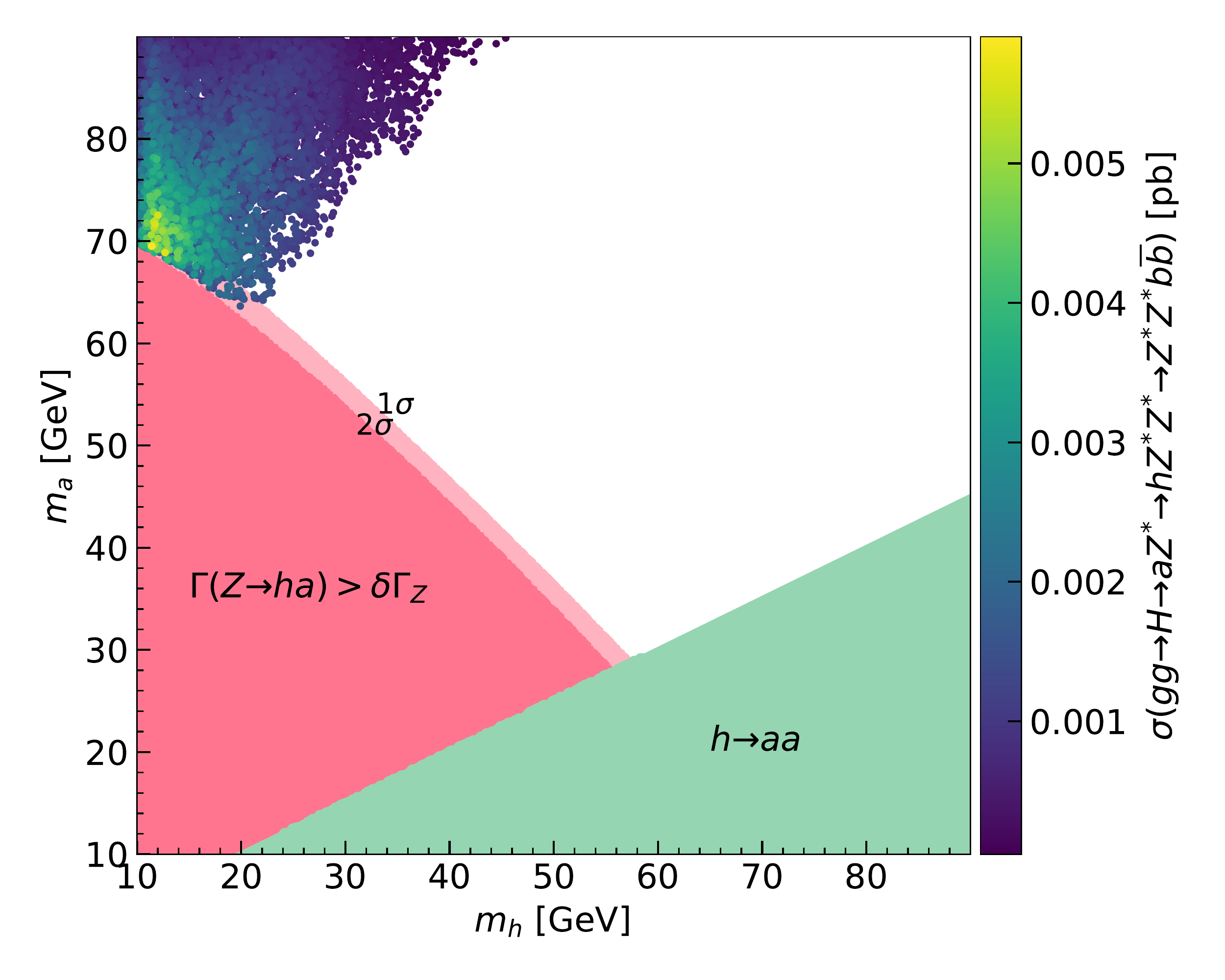}
	\caption{$m_h$ and $m_a$ vs. ${\rm BR}(h \to b\overline{b})$ (left) and $\sigma(gg \to H \to aZ^{*} \to h Z^{*}Z^{*} \to Z^{*}Z^{*}b\overline{b})$ (right) at 95\% C.L. in the 2HDM Type-I.}
	\label{fig6}
\end{figure}

We show in Fig.~\ref{fig6} the  $gg \to H \to aZ^{*} \to h Z^{*}Z^{*}$ cross section, where $h \to b\overline{b}$. The process could yield a cross section of 0.006 pb. In the right panel of Fig.~\ref{fig6} we show the BR of $h \to b\overline{b}$ in this  region of the 2HDM Type-I parameter space. Obviously, the decay width of $h$ is dominated by the decay mode $ h \to b\overline{b}$. Thus, in what follows, we focus on the case where $h$ decays to $b\overline{b}$ and $Z^{(*)}Z^{(*)} \to \mu^+\mu^- jj$. Such a scenario could be an alternative  channel to search for light Higgses at Run 3 and the HL-LHC.

\section{Signal vs. Background Analysis}
\label{section4}
We describe here the toolbox used to generate and analyse MC events. \texttt{MadGraph-v.9.2.5}~\cite{Alwall:2014hca} is used to generate parton level configurations of both signal and background processes\footnote{Background and signal events are generated at LO. Higher order corrections are quantified through  $K$-factors. \textcolor{black}{The NLO QCD correction to top pair production in association with 2 jets computed at the LHC is about  $-27\%$~\cite{Bevilacqua:2011aa}, which we adopt here. The NLO corrections to $gg \to H$ are very large {\tiny {\tiny }}, about a facor of 2, due to the contributions from $gg$ pairs to QCD radiation, whereas ${K}_{\text{NNLO/NLO}}\approx$ is much smaller than ${K}_{\text{NLO/LO}}$, signifying a convergence of the QCD expansion, so we renormalise the signal to the NNLO rates through the {$K$-factor} $K_{\rm NNLO/LO} = \sigma^{\rm NNLO}/\sigma^{\rm LO} \sim 2.6-2.7$.}}. The events are passed then to \texttt{PYTHIA8}~\cite{Sjostrand:2006za} to simulate parton showering, hadronization and decays. Finally, we use \texttt{Delphes-3.5.0}~\cite{deFavereau:2013fsa} with the standard CMS card\footnote{It adopts the anti-$k_T$ algorithm to cluster final particles into jets, with jet parameter $\Delta R$ = 0.5 and $p_{T,j}^{\rm min} = 20$ GeV (for both light- and $b$-quark jets).} to perform  detector simulation. We resort to \texttt{MadAnalysis}~\cite{Conte:2012fm} to apply cuts and to conduct the analysis. Background processes with dominant contributions are top pair production in association with 2 Initial State Radiation (ISR) jets\footnote{In our study, we focus mainly on $ggt\overline{t}$ which is vastly dominant over $gq\overline{q}t\overline{t}$ and $q\overline{q}t\overline{t}$.} and $ZZ$ production with additional $b\overline{b}$ quarks.  We show in Tab.~\ref{tab-3} the corresponding cross sections at $\sqrt{s}=13$ TeV for the LHC energy.  We have generated MC samples of $O(10^6)$ events. Unsurprisingly, the irreducible background $pp \rightarrow Z^{(*)}Z^{(*)}b\overline{b} \to b\overline{b}jj\mu^+ \mu^-$ (from both QCD and EW interactions) is negligible whereas $pp \rightarrow gg t\overline{t} \rightarrow gg\mu^+\mu^- jj b\overline{b} \nu_\mu \overline{\nu}_\mu$ is large. 
\begin{table}[h!]
	\begin{center}
		\resizebox{0.47\textwidth}{!}{
			\begin{tabular}{||c|c||} \hline\hline
				Background  &   Cross section (pb)  \\   
				\hline \hline
				$pp \rightarrow ZZb\overline{b}_{\rm QCD} \rightarrow \mu^+\mu^- jj b\overline{b}$&~  $9.27\times 10^{-3}\pm 2.4\times 10^{-5}$ \\
				\hline 
				$pp \rightarrow ZZb\overline{b}_{\rm QED} \rightarrow \mu^+\mu^- jj b\overline{b}$&~  $2.42\times 10^{-4}\pm 5.5\times10^{-7}$  \\
				\hline 
				$pp \rightarrow gg t\overline{t} \rightarrow gg\mu^+\mu^- jj b\overline{b} \nu_\mu \overline{\nu}_\mu$ &~ $2.92\pm 0.008$\\
				\hline
		\end{tabular}}	
	\end{center}
	\caption {The parton level cross sections of the background processes at LO.}
	\label{tab-3}
\end{table}

\begin{figure}[h!]
	\centering
		\resizebox{0.4\textwidth}{!}{\includegraphics{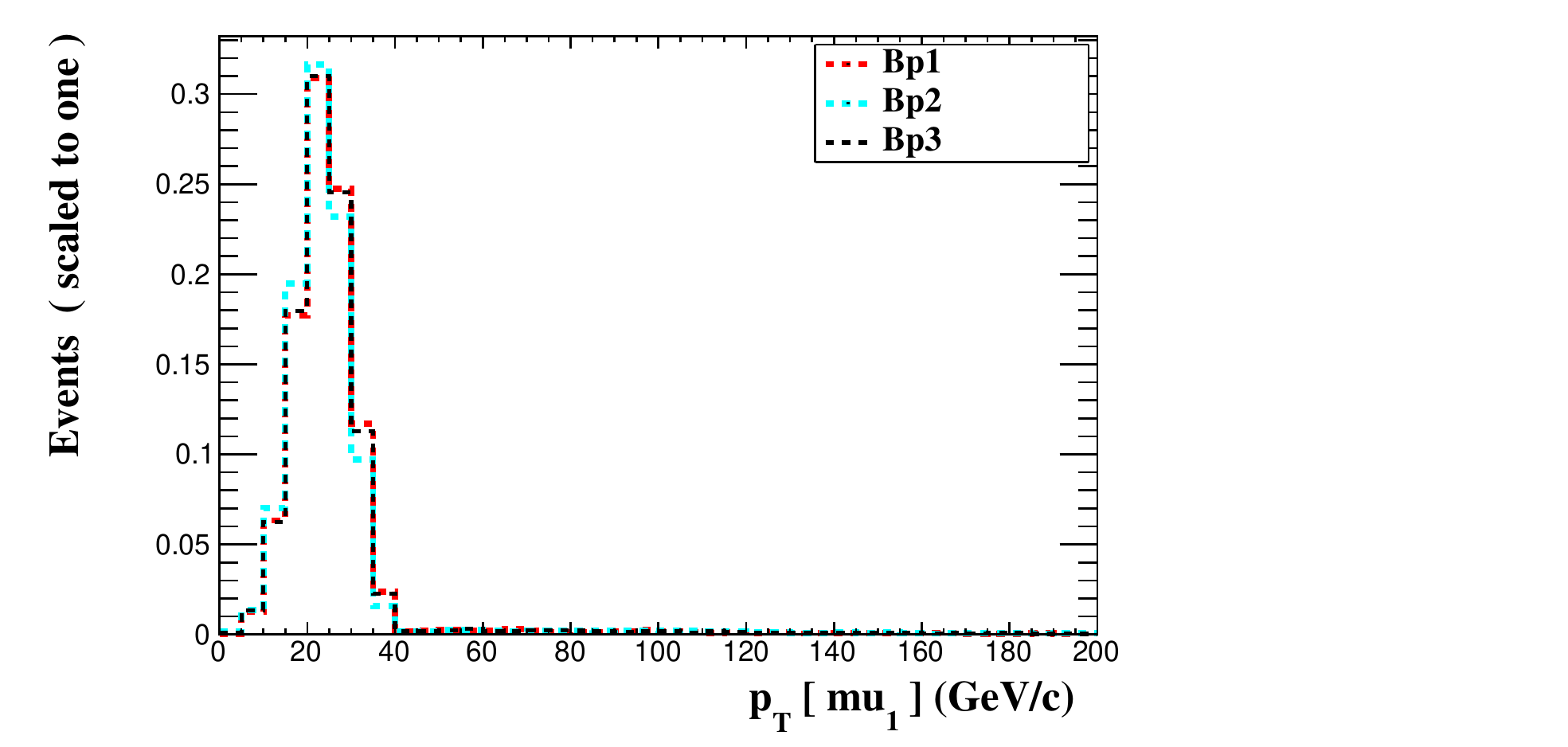}}
		\resizebox{0.4\textwidth}{!}{\includegraphics{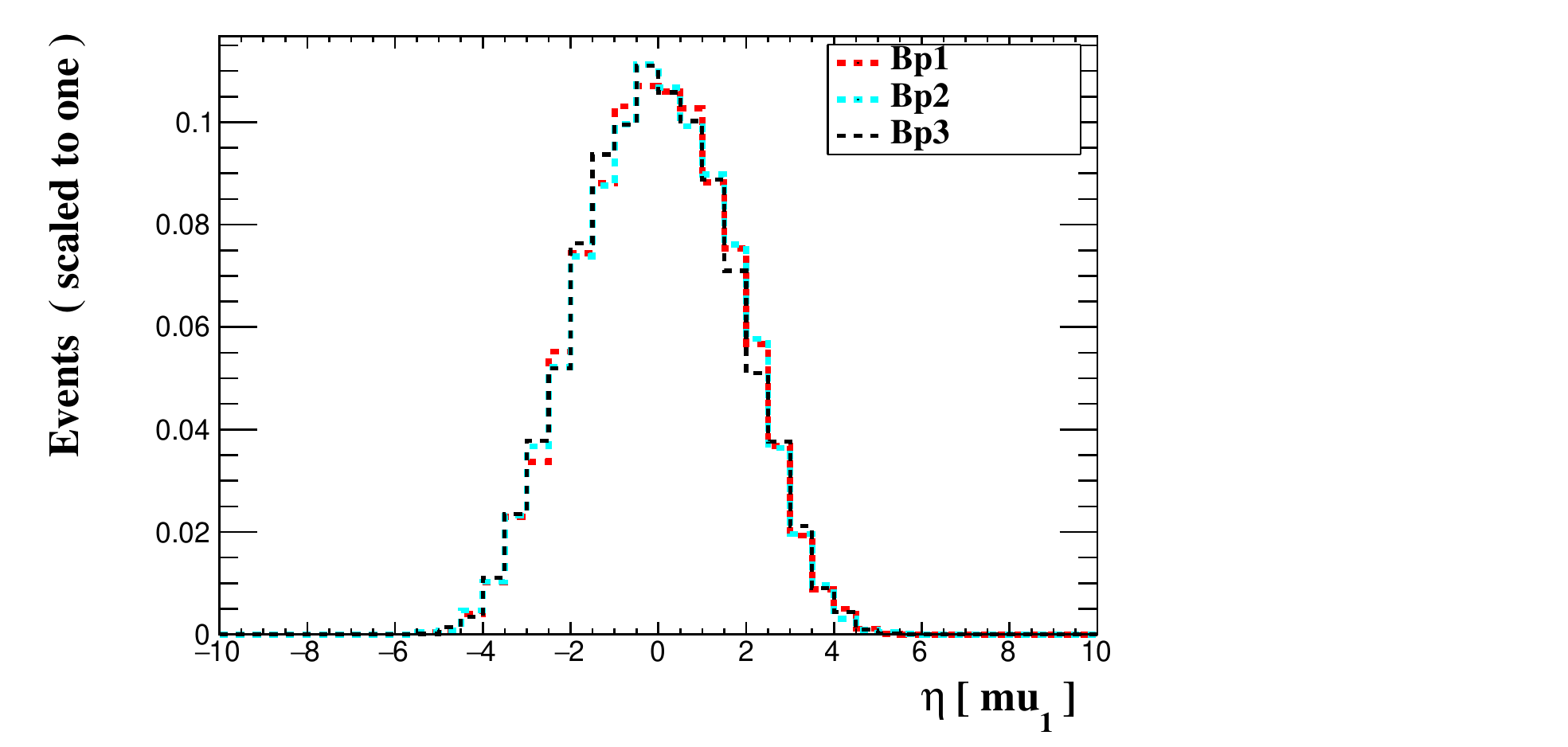}}
	\caption{The transverse momentum (left) and pseudorapidity (right) of the hardest muon for the signal (all BPs).}	
	\label{fig4}
	\centering
	\resizebox{0.4\textwidth}{!}{\includegraphics{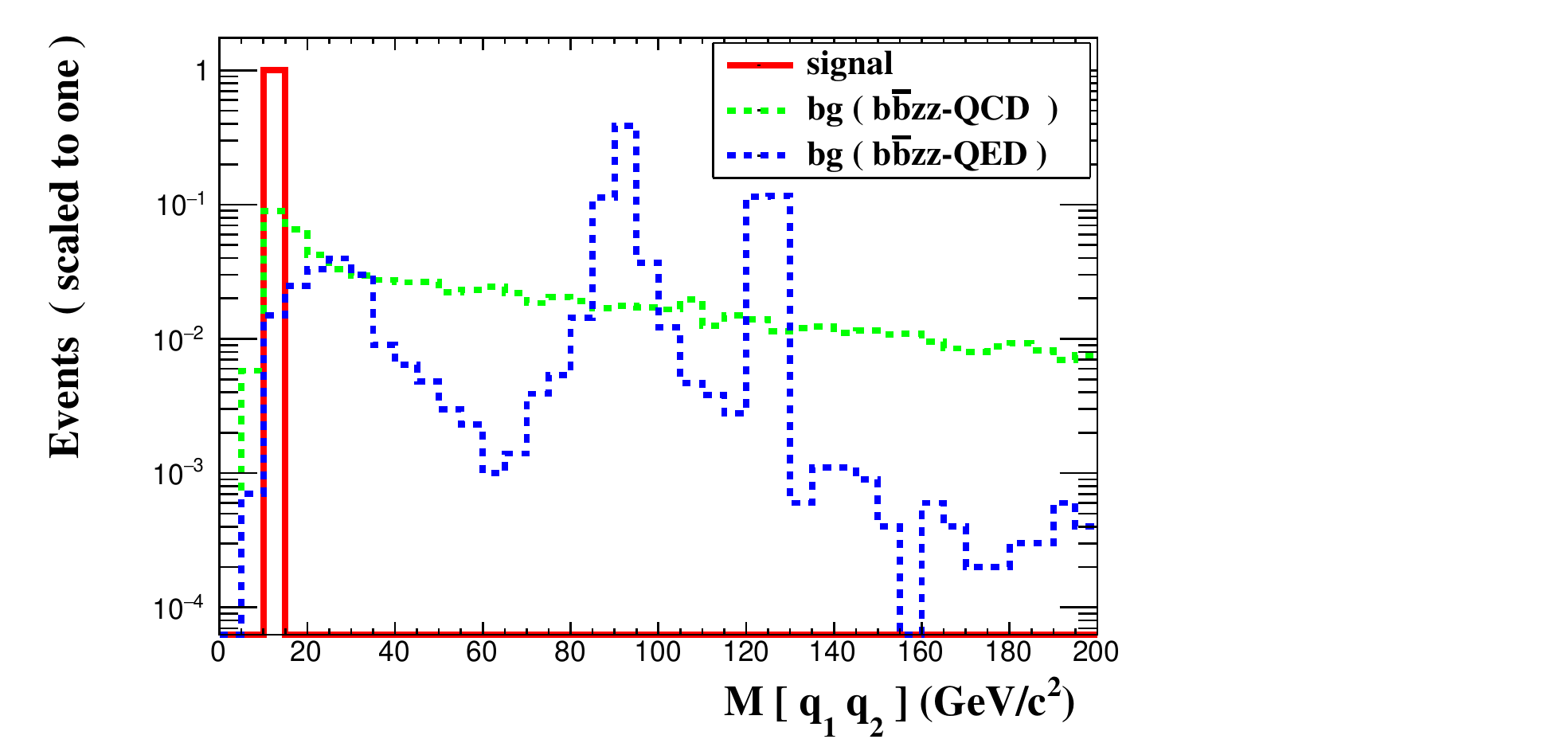}}
	\resizebox{0.4\textwidth}{!}{\includegraphics{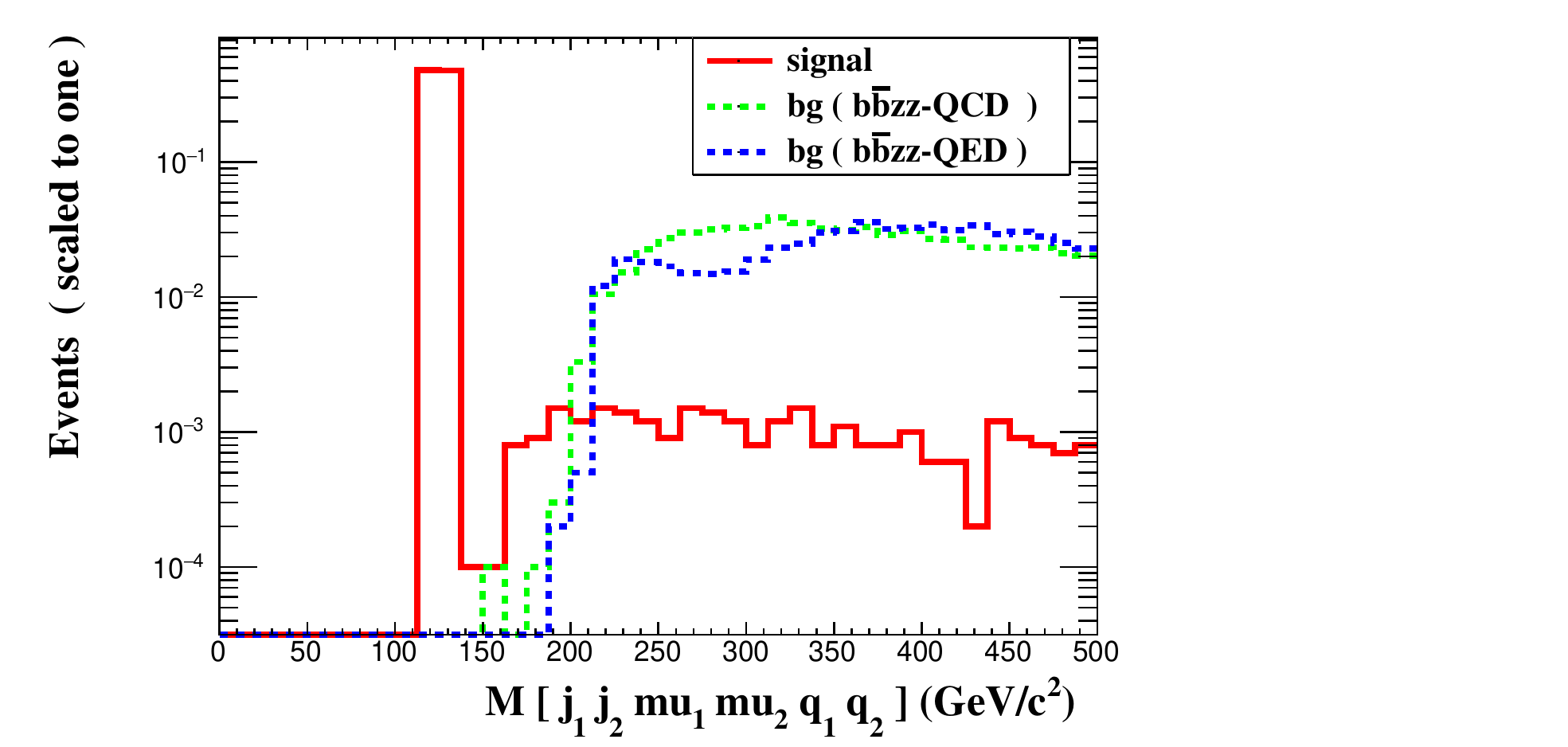}}
	\caption{The invariant mass of the $b\bar b$ (left) and $\mu^+ \mu^- ~jj~b\overline{b}$ (right) system for the signal (BP1) and the two irreducible backgrounds at parton level.}	
	\label{figg5}
\end{figure}

We considered a few BPs for the signal given by $gg\to H \to aZ^{*}
\to h Z^{*}Z^{*} \to \mu^+ \mu^- ~jj~b\overline{b}$ to perform the MC
simulation. The input parameters of each BP are given in
Tab.~\ref{tab-4}. Note that the light Higgs width, $\Gamma(h)$, is not
small enough to lead to a large lifetime and hence, long-lived
particles producing displaced vertices inside the detector. The proper
decay length $c\tau_0$ is in fact only a tiny fraction of
micrometers\footnote{The proper decay length $c\tau_0$ falls within
the range from $0.06\mu m$ to $0.19\mu m$, where $\tau_0$ is the light
Higgs lifetime at rest~\cite{Accomando:2016rpc}.}.  The different
kinematic distributions at parton level in Fig.~\ref{fig4} show that
the requirement of central pseudorapidity of the muons is generally
satisfied however the $p_T$ of these can be rather small, so that we
will invoke the CMS di-muon trigger of Ref.~\cite{CMS:2021yvr},
whereby the threshold is 17 GeV~\cite{CMS:2021yvr} for the muon with
highest $p_T$ and 8 GeV for the other. Fig.~\ref{figg5} shows the
invariant mass distributions of the two $b$-jets, $m_{b\overline{b}}$,
and that of the full final state, $m_{jj \mu^+ \mu^- b\overline{b}}$,
for the signal and the irreducible background processes at parton level, noting that
$m_{b\overline{b}}$ is close to light Higgs mass $m_h$ and $m_{jj
  \mu^+ \mu^- b\overline{b}}$ is close to SM-like Higgs mass $m_H$
(for the signal, unlike the irreducible backgrounds). We will clearly
leverage these underlying partonic shapes in our detector level
analysis, to which we proceed next, in the presence of the following
sequence of acceptance cuts:
\begin{equation*}
p_T^{j,~b} > 20~\text{GeV},~p_T^l > 10~\text{GeV},~|\eta(l,b)|< 2.5,~|\eta(j)|< 5.0,~\Delta R > 0.4
\end{equation*}
\begin{table}[!h]
	\begin{center}
		\resizebox{0.95\textwidth}{!}{
			\begin{tabular}{c|c|c|c|c|c|c|c|c|c|c|c|c|} \hline\hline
				BPs  & ~$m_h$ &~  $m_H$ & ~ $m_a$ &~ $m_{H^\pm}$ &~$\sin(\beta-\alpha)$ & ~$\tan \beta$ & $\Gamma(h)$ &$\Gamma(H)$ &$\Gamma(A)$ &$\Gamma(H^\pm)$ & $\Gamma(Z \to h a)$ & $\sigma$ (pb)\\ 
				\hline
				BP1&~ 15.37 &~ 125.00 &~72.21&~120.99&~ $-0.19$ &~ 8.55& $3.11\times 10^{-9}$ & $4.4\times 10^{-3}$& $1.028\times 10^{-4}$ & $7.88 \times 10^{-2}$ & 0.00083 & $3.28\times 10^{-4}$ \\
				\hline  
				BP2&~ 12.56 &~ 125.00 &~74.12&~113.93&~ $-0.16$ &~ 5.97& $1.01\times 10^{-9}$ & $4.4\times 10^{-3}$& $1.60\times 10^{-4}$ & $4.752 \times 10^{-2}$ & 0.000968 & $4.11\times 10^{-4}$\\
				\hline
				BP3&~ 11.64 &~ 125.00 &~73.03&~104.56&~ $-0.19$ &~ 5.09& $3.13\times 10^{-9}$ & $4.49\times 10^{-3}$& $1.644\times 10^{-4}$ & $3.96 \times 10^{-2}$ & 0.00164 &  $ 4.73\times 10^{-4}$\\
				\hline 
		\end{tabular}}
	\end{center}
	\caption {Selected BPs with parton level cross section and other observables at {{LO}}. (All masses and widths are in GeV.)}
	\label{tab-4}
\end{table}

\begin{figure*}[!h]
	\centering
	\resizebox{0.4\textwidth}{!}{
		\includegraphics{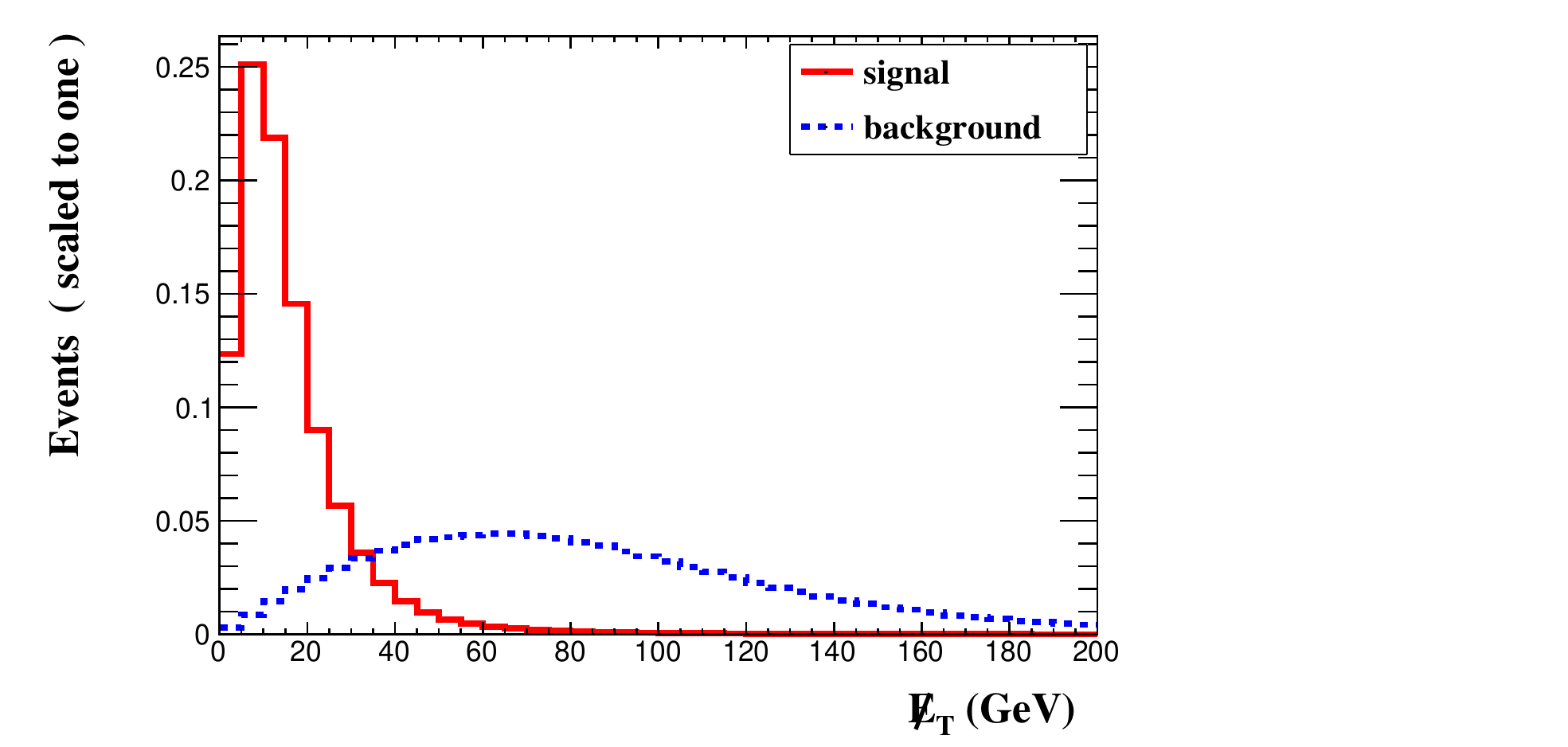}}
	\resizebox{0.4\textwidth}{!}{
		\includegraphics{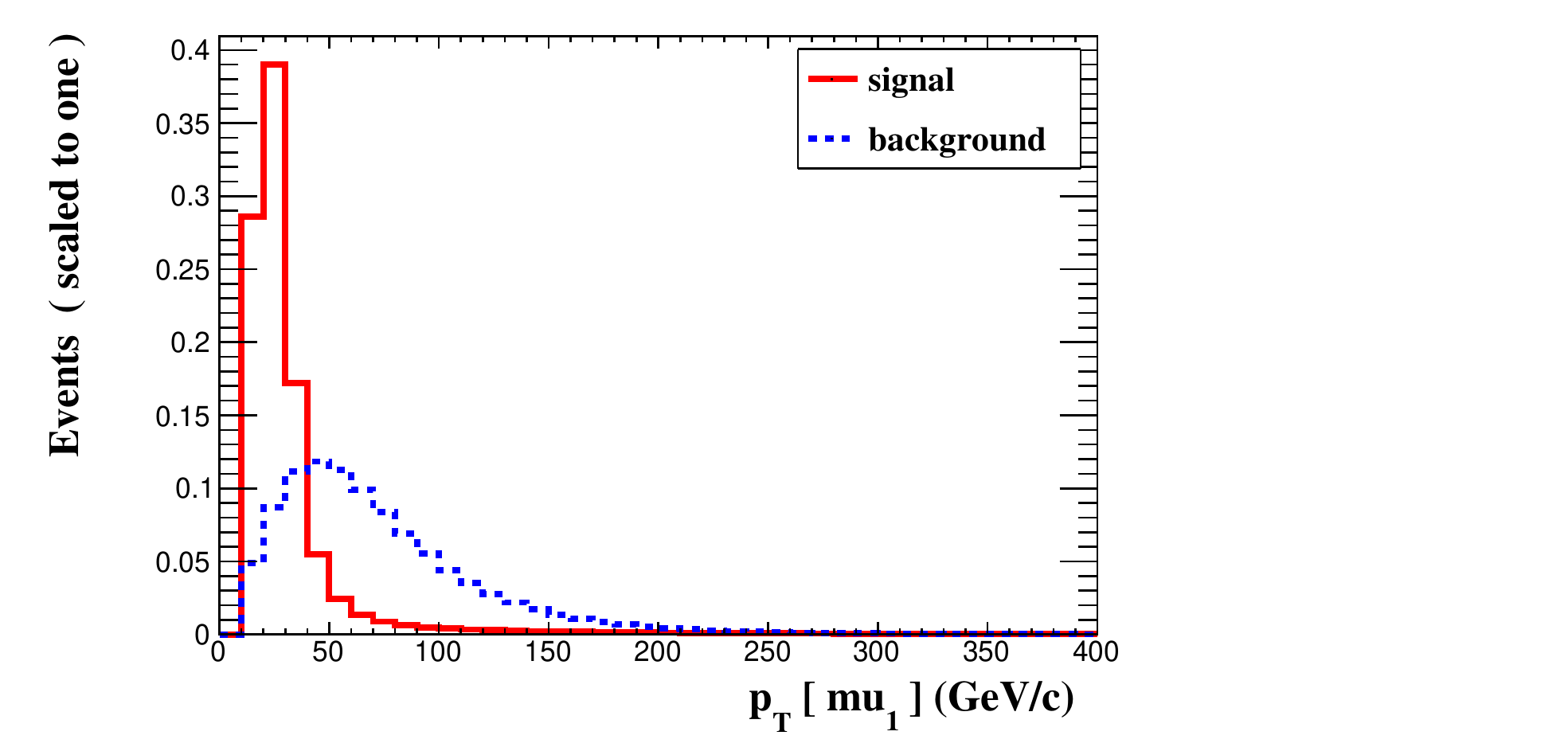}}
	\resizebox{0.4\textwidth}{!}{
		\includegraphics{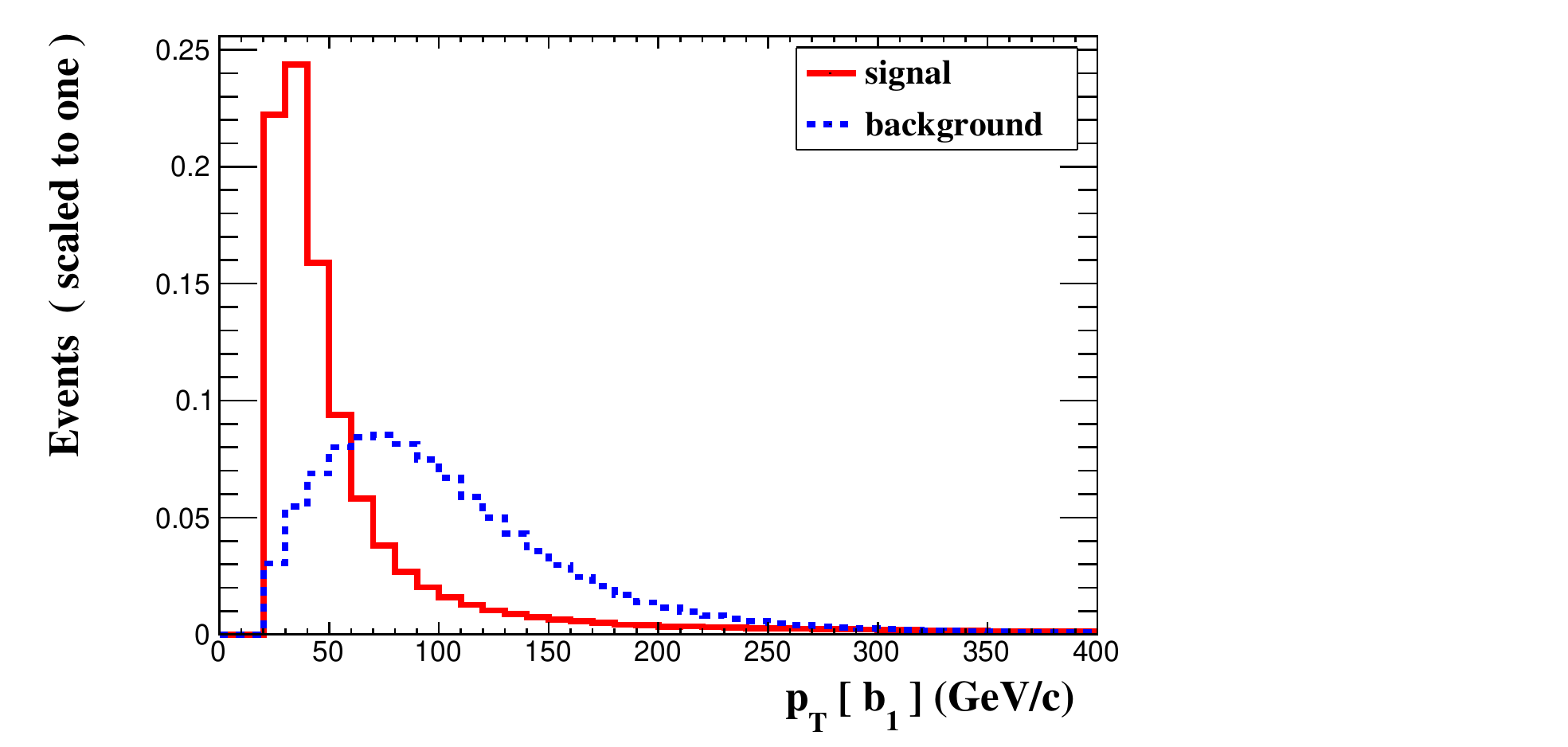}}
	\resizebox{0.4\textwidth}{!}{
		\includegraphics{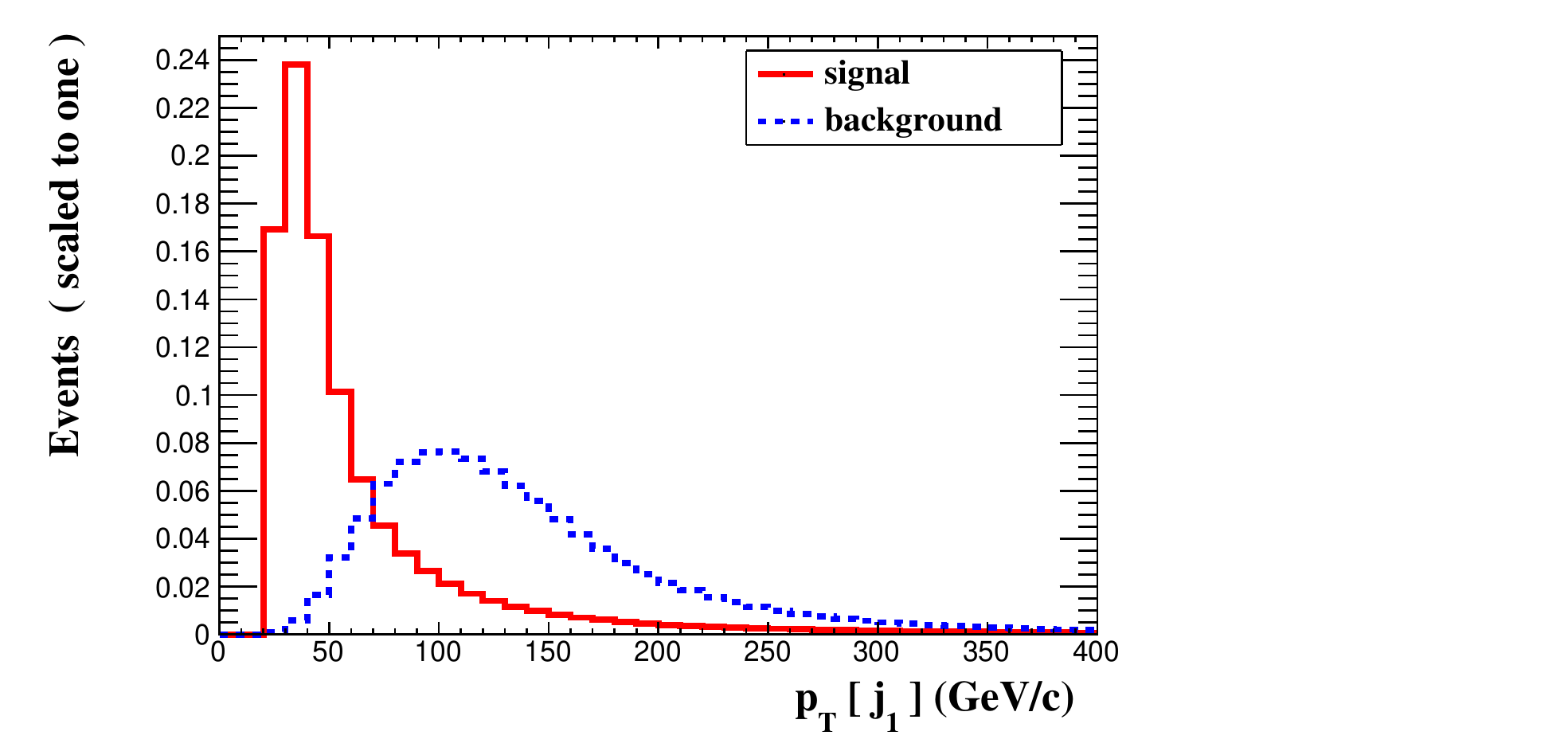}}
	\caption{The $\eslash_{\rm T}$ and highest $p_T$ distributions for muons, $b$- and light-jets (clockwise) for signal (BP3) and background (blue)  ($ggt\bar t$) at detector level.}
	\label{figg6}
\end{figure*}

\begin{figure*}[!h]
	\centering
	\centering
	\resizebox{0.4\textwidth}{!}{
		\includegraphics{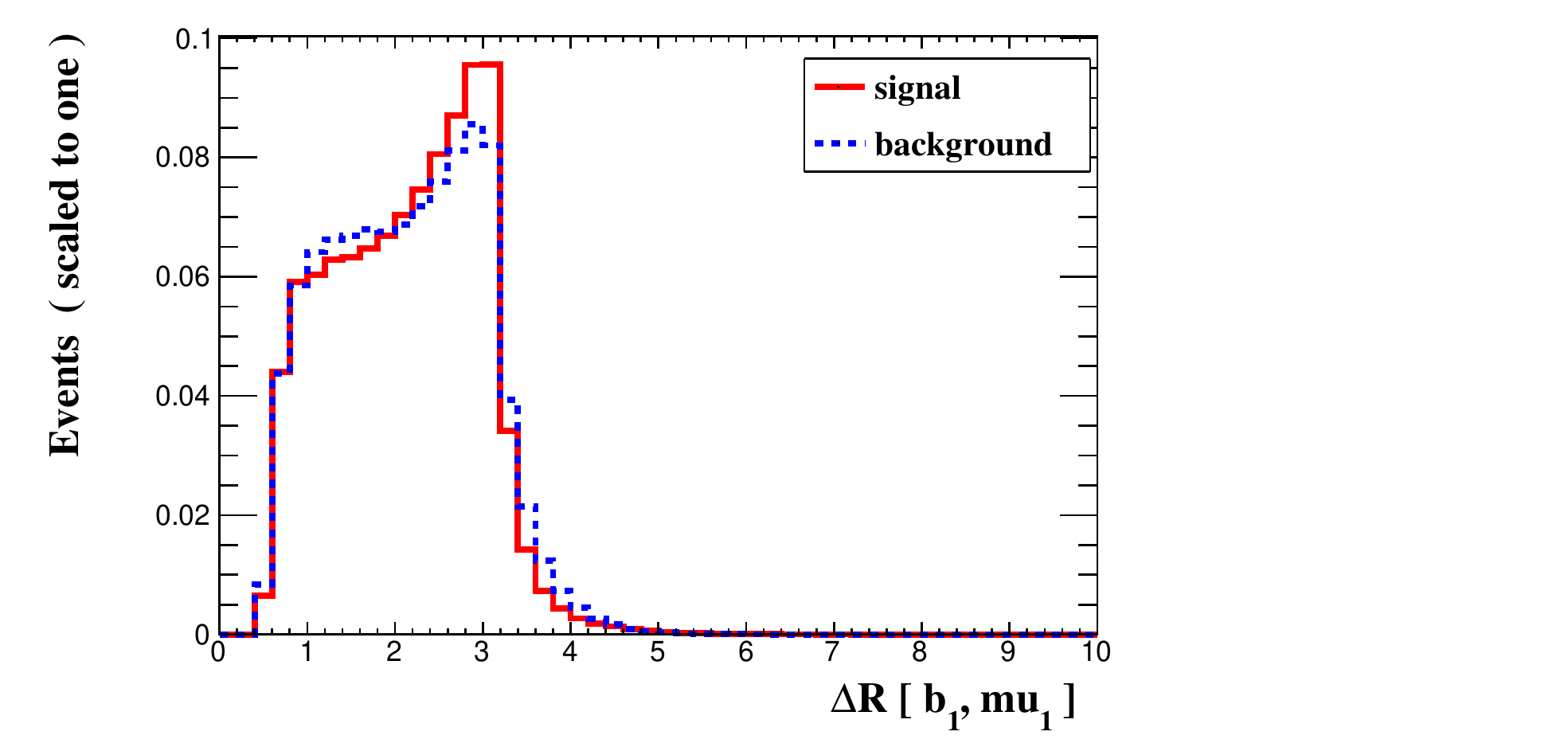}} 
	\resizebox{0.4\textwidth}{!}{
		\includegraphics{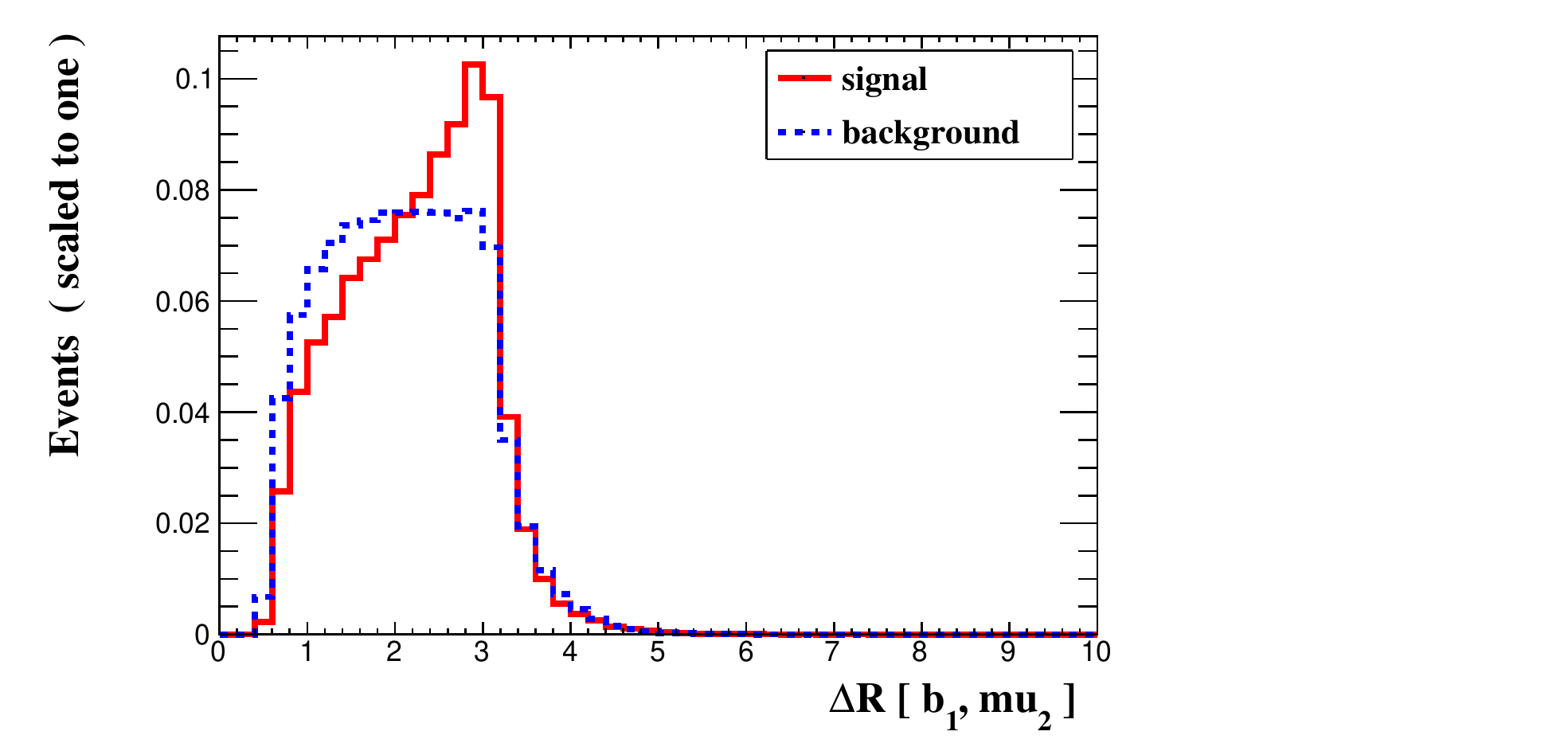}}\\
	\resizebox{0.4\textwidth}{!}{
		\includegraphics{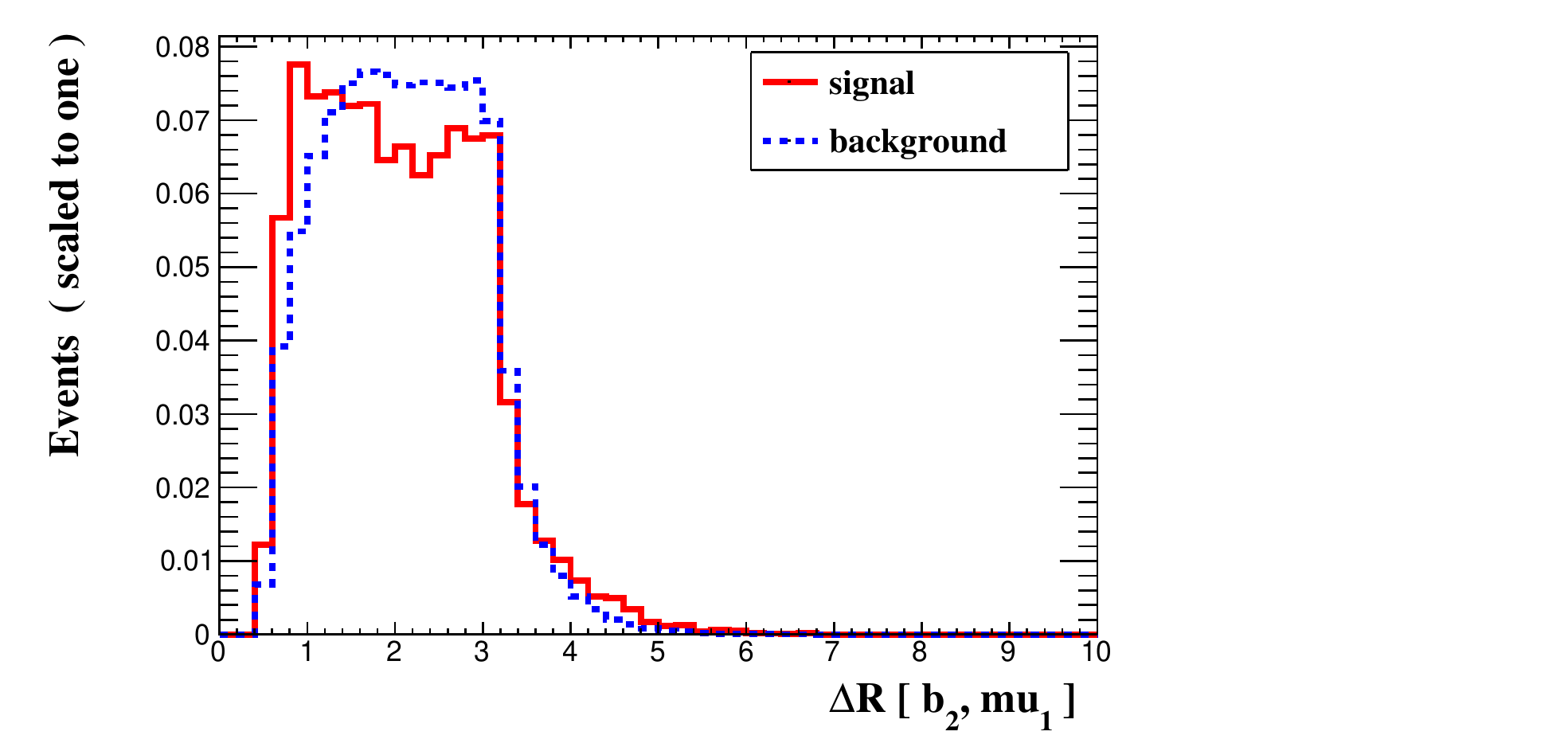}}
	\resizebox{0.4\textwidth}{!}{
		\includegraphics{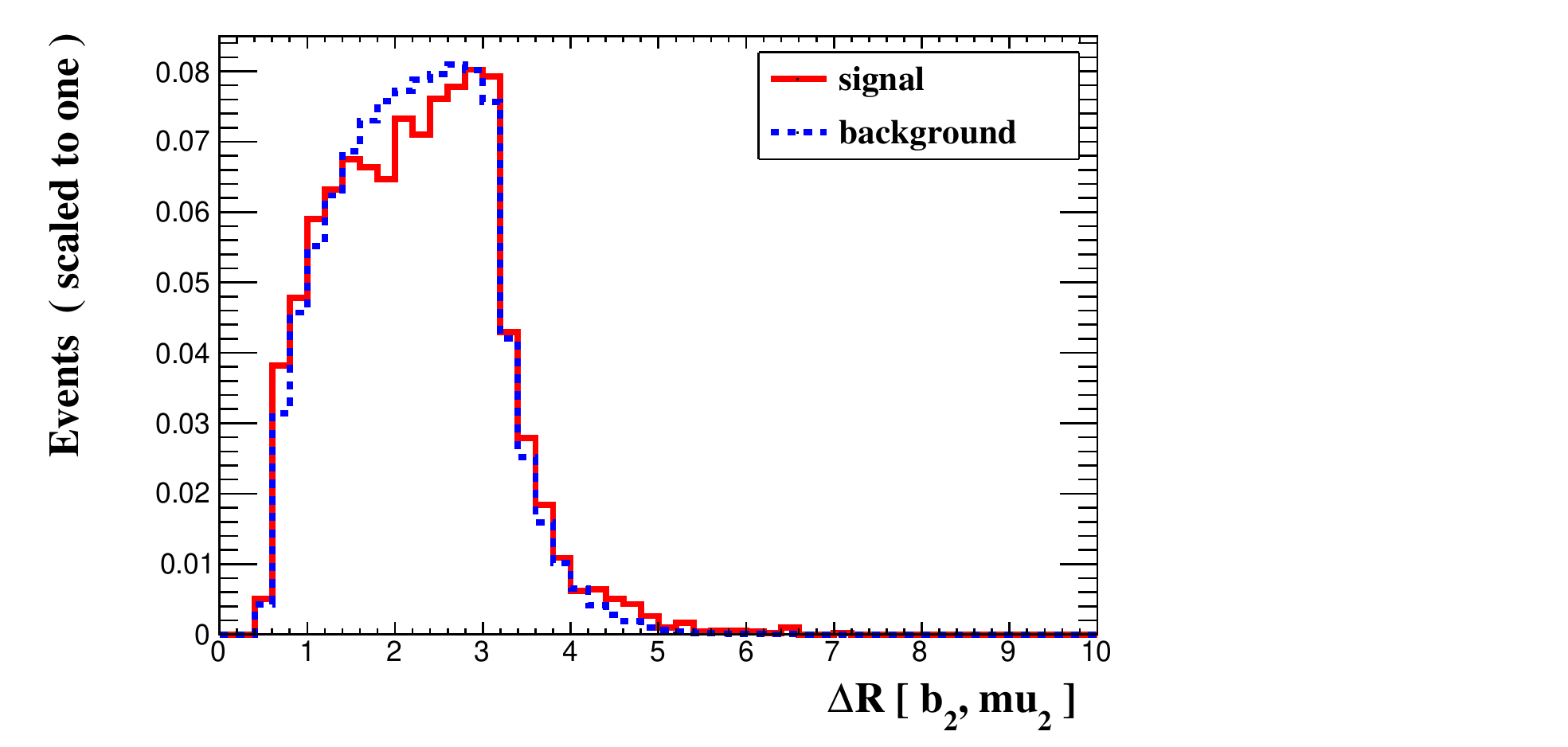}}
	\caption{$\Delta R$ distributions between the two ($p_T$ ordered) $b$-jets and muons, from hardest to softest (clockwise) for signal (red) (BP3) and background (blue) at detector level.}
	\label{figg7}
\end{figure*}

We show in Fig.~\ref{figg6} the distributions of the missing transverse energy ($\eslash_{\rm T}$) and the highest $p_T$'s of $b$-jets, light-jets and muons for signal and background at detector level. (As mentioned previously, the irreducible background stemming from $ZZ b\overline{b}$ processes is negligible, so we have not emulated these at detector level.) The $\eslash_{\rm T}$ distribution from simulated samples of background events is mainly from di-leptonic decay of $ggt\overline{t}$, i.e., with $t \overline{t}\to W^+bW^-\overline{b} \to (\mu^+ \nu_\mu b) (\mu^- \overline{\nu}_\mu \overline{b})$. An interesting observation is the $\eslash_{\rm T}$ in the signal, which is given by $H \to aZ^{*} \to h Z^{*}Z^{*} \to \mu^+ \mu^- ~jj~b\overline{b}$ events with semi-leptonic $b$-meson decays (alongside detector effects). Furthermore, Fig.~\ref{figg7} illustrates the different angular separations between $b$-quarks and muons for signal and background, where one can read that background has only a minimal component with muons coming from semi-leptonic $b$-meson decays (as intimated). 

\begin{figure*}[h!]
	\centering
	\resizebox{0.32\textwidth}{!}{
	\includegraphics{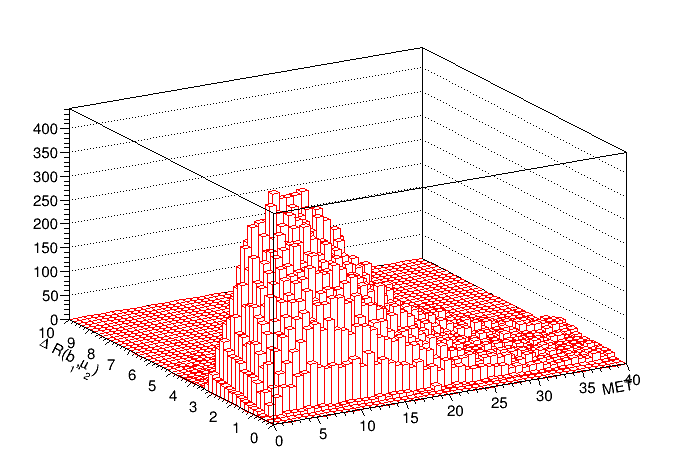}}
\resizebox{0.32\textwidth}{!}{
	\includegraphics{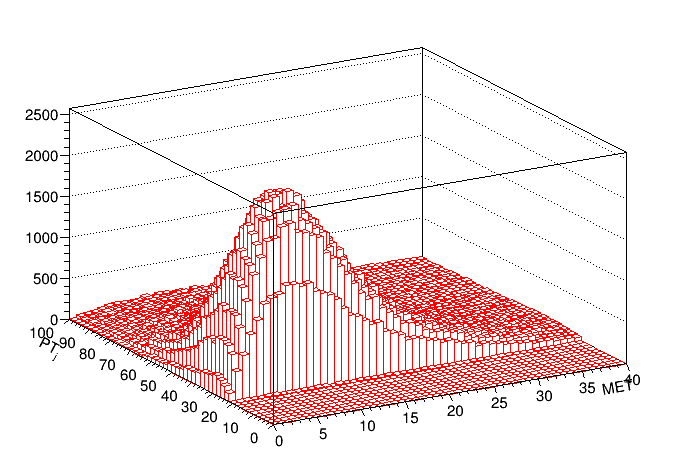}}
\resizebox{0.32\textwidth}{!}{
	\includegraphics{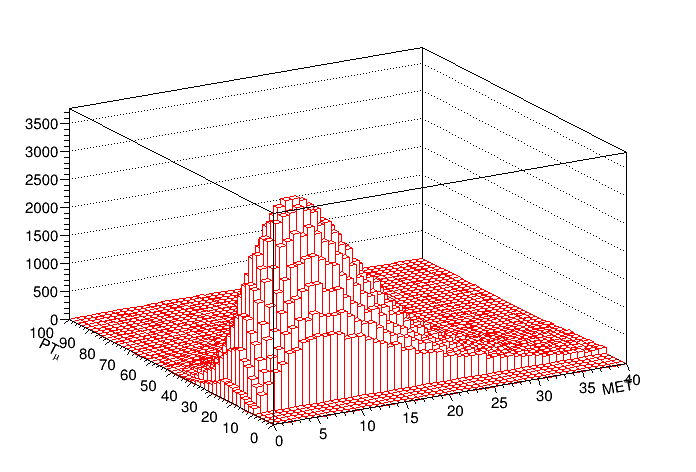}}
\caption{Correlation between $\Delta R(b_1,\mu_2)$ vs. $\eslash_{\rm T}$ (left panel), $p_T^j$ vs. $\eslash_{\rm T}$ (middle panel) and $p_T^\mu$ vs. $\eslash_{\rm T}$ (right panel) for the signal (BP3) at detector level.}
		\label{figg8}
	\resizebox{0.32\textwidth}{!}{
	\includegraphics{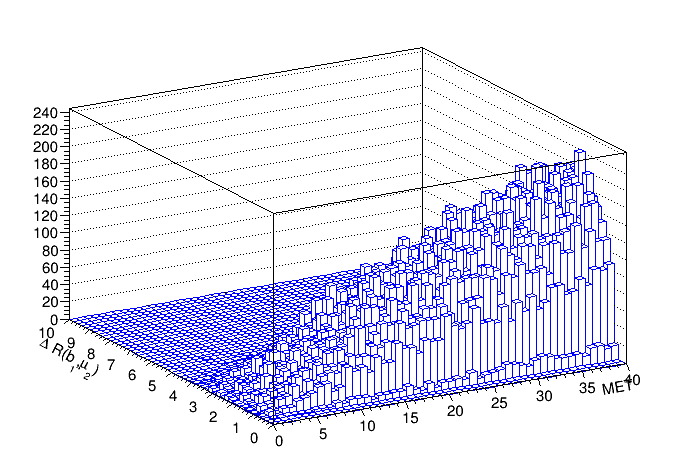}}
\resizebox{0.32\textwidth}{!}{
	\includegraphics{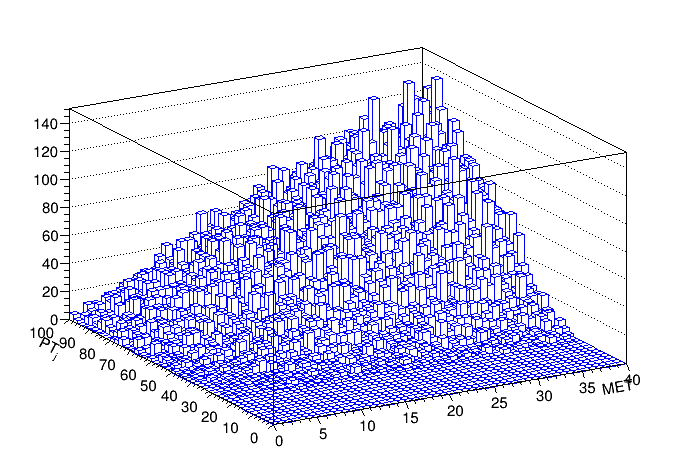}}
\resizebox{0.32\textwidth}{!}{
	\includegraphics{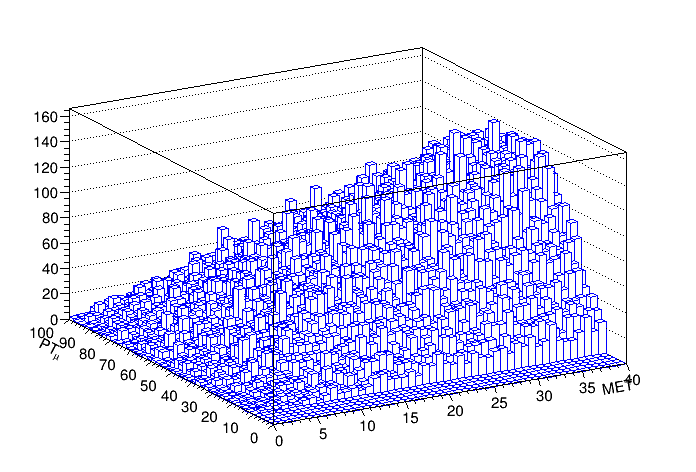}} \\
\caption{Correlation between $\Delta R(b_1,\mu_2)$ vs. $\eslash_{\rm T}$ (left panel), $p_T^j$ vs. $\eslash_{\rm T}$ (middle panel) and $p_T^\mu$ vs. $\eslash_{\rm T}$ (right panel) for the background ($ggt\bar t$) at detector level.}
	\label{figg9}
\end{figure*}

To enhance the signals and  suppress the background from $ggt\overline{t}$, we have adopted several kinematic cuts, which   choice  is based on comparing different distributions of the signal and background processes at the detector level. Specifically, this has been done through 2D distributions correlating the
missing transverse momentum to a series of kinematic variables pertaining to some of the visible objects in the final state as illustrated in Figs.~\ref{figg8} and \ref{figg9} for signal and background, respectively.  An interesting observation is that the signal and background distributions are anti-correlated. In fact, it is clear from the left panel of Figs.~\ref{figg8} and \ref{figg9} that forcing the missing transverse energy to be below 25 GeV will strongly favour the signal over the background. The middle and right panel show  that selecting events with $p_T^{j} < 75$ GeV and $p_T^{\mu} < 40$ GeV  would also enhance the signal significance. Through similar reasoning, further cuts were required on $m_{\mu\mu}$, $\Delta R (b_i, \mu_j)$, $\Delta R (\mu_1, \mu_2)$,  $\Delta R (j_1, j_2)$ and $\Delta R (b_1, b_2)$. These selection cuts were finally chosen as follows: $m_{\mu\mu} < 40~\text{GeV}$, $\Delta R (b_i, \mu_j)<2.5$, $\Delta R (\mu_1, \mu_2)<2.5$,  $\Delta R (j_1, j_2)<2.5$ and $\Delta R (b_1, b_2)<2.5$.

\begin{figure*}[!h]
	\centering
		\includegraphics[width=0.45\textwidth]{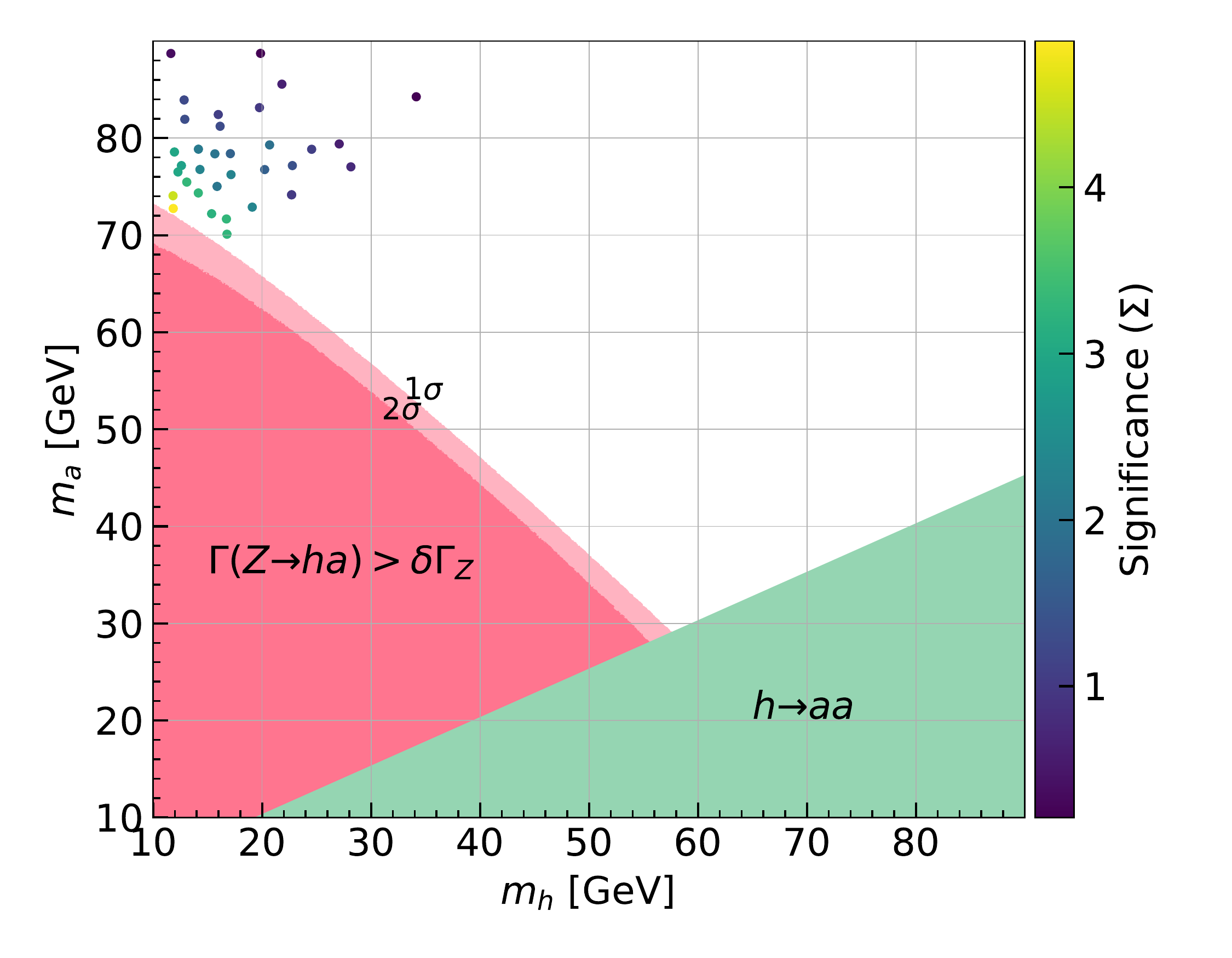}
	\includegraphics[width=0.45\textwidth]{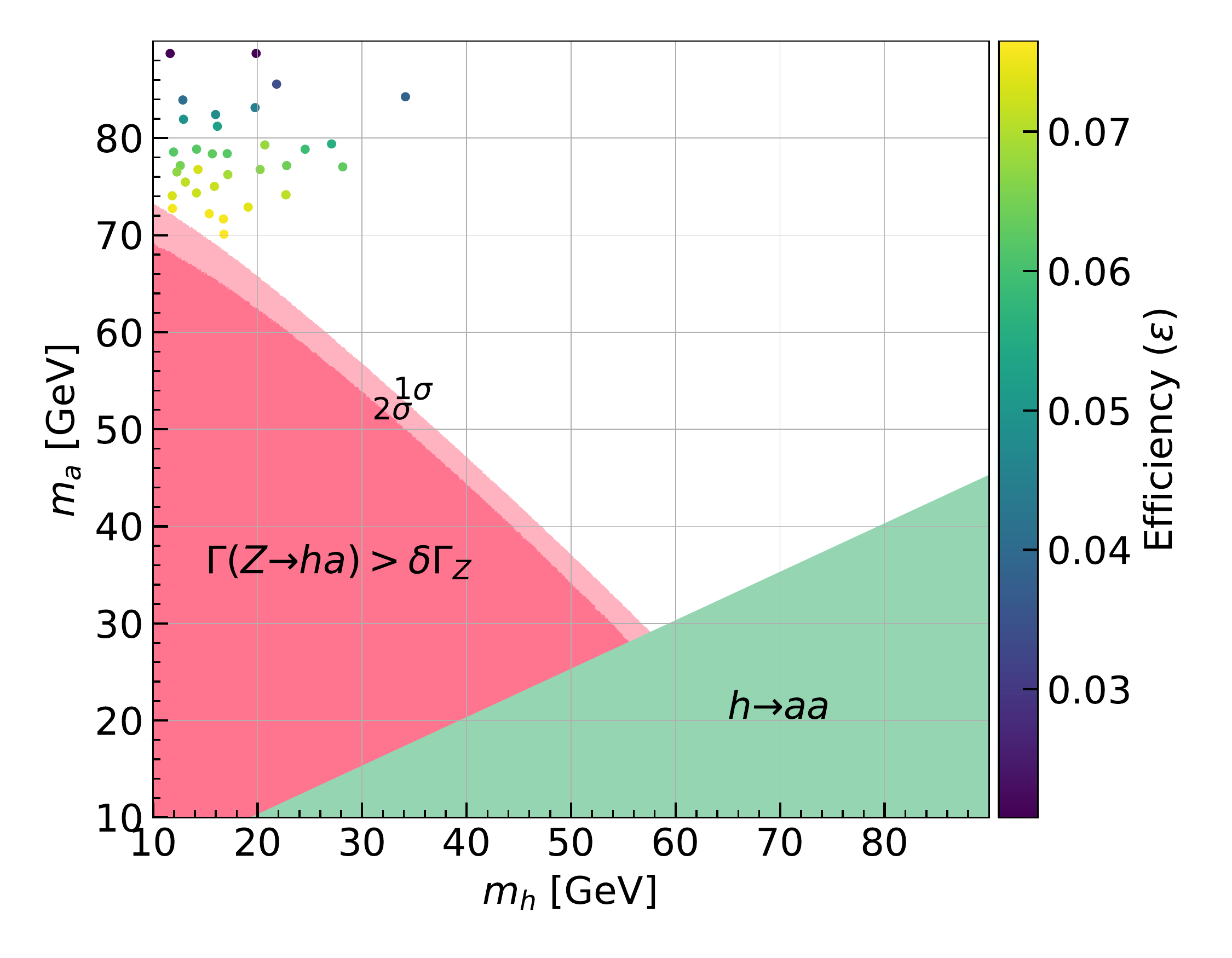} \\
	\caption{Significance  (left) and efficiency (right) of each BP produced in our analysis over the ($m_h, m_a$) projection of the 2HDM Type-I  parameter space, after the full cutflow described in the text. }
	\label{figg0}
\end{figure*}

We have then computed  the significance  (for $\sqrt{s}=13$ TeV and $\mathcal{L} = 300~ \text{fb}^{-1}$), defined as $\Sigma=\frac{S}{\sqrt{B}}$, where $S$($B$) is the signal(background) yield after the discussed cutflow,  for not only our three initial BPs (whose $\Sigma$ rates are 3.20, 4.12 and 4.80 for BP1, BP2 and BP3, respectively), but also those appearing in 
 Tab.~\ref{tab5}. We have done so in order to be able to map the 2HDM Type-I parameter space in
detail, so as to acquire a sense of the true portion of it that can be tested by forthcoming experiments. Note that we have kept the same cutflow already illustrated for all such new BPs too. Also, it is at this stage that we take into account the aforementioned QCD $K$-factors for both signal and background. 
The applied kinematic cuts can greatly reduce the $ggt\bar t$ background, as their efficiency is here 
$\epsilon = \frac{\text{cross section after cuts}}{\text{cross section before cuts}}=0.000056$, i.e., much lower than the values associated to all our   
 BPs. Many of the latter can have a significance larger than 3 and up to nearly 5, for Run 3 energy and luminosity. To observe their distribution over the ($m_h, m_a$) plane, we have finally produced Fig.~\ref{figg0}, indeed, assuming  $\sqrt{s}=13$ TeV and $\mathcal{L} = 300~ \text{fb}^{-1}$, where both $\Sigma$ and $\epsilon$ are mapped. Hence, at Run 3, we can conclude that a substantial portion of the 2HDM Type-I parameter space can offer some evidence (and even near discovery)
of the signal we have pursued. Furthermore, we notice  that a larger efficiency can be obtained for small $m_a$: this is because the loss of efficiency with $b$-tagging is over-compensated by a simultaneous higher efficiency for both $j$- and $\mu$-tagging. Needless to say, at the HL-LHC, where $\mathcal{L} = 3000~ \text{fb}^{-1}$,
most of the sampled parameter space of the 2HDM Type-I would be discoverable.

\begin{table}[h!]
	\begin{center}
		\scalebox{0.8}{
			\begin{tabular}{||c|c|c|c|c|c|c|c||} \hline\hline
				BP & $m_h$ (GeV)  & $m_a$ (GeV) & $\sigma$ (pb) & $K$-factor & No. of events    & Significance $\Sigma$ & Efficiency $\epsilon$ \\  
				\hline
				BP4 & 11.85 & 72.75 & $4.82\times 10^{-4}$ & 2.68  & 10.97   & 4.88 &  0.0758  \\
				\hline
				BP5 & 17.15 & 76.24 & $2.54\times 10^{-4}$ &  2.63 &  5.254  & 2.29  & 0.0689  \\
				\hline
				BP6 & 24.55 & 78.85 & $1.39\times 10^{-4}$ & 2.62  &  2.487 &  1.08   & 0.059   \\
				\hline
				BP7 & 15.98 & 82.43 & $1.705\times10^{-4}$ & 2.63  & 2.465  & 1.07  & 0.048  \\
				\hline
				BP8 & 34.15 & 84.26 & $4.48\times 10^{-5}$ &  2.60  & 0.512  & 0.22  & 0.038  \\
				\hline
				BP9 & 20.69 & 79.30 & $2.30\times 10^{-4}$ &  2.63  & 4.141  & 1.93 & 0.068   \\
				\hline
				BP10 & 16.73 & 71.67 & $3.31\times 10^{-4}$ &  2.63  & 7.539  & 3.29  & 0.0758  \\
				\hline
				BP11 & 16.78 & 69.25 & $3.247\times 10^{-4}$ &  2.63  & 7.460   & 3.26  & 0.0765  \\
				\hline
				BP12 & 21.82 & 85.56 & $1.42\times 10^{-4}$ &  2.64  & 1.455   & 0.63  & 0.034  \\
				\hline
				BP13 & 22.78 & 77.17 & $1.629\times 10^{-4}$ &  2.62  & 3.144   & 1.369 & 0.0643  \\
				\hline		 
				BP14 & 17.09 & 78.40 & $2.038\times 10^{-4}$ &  2.63  & 3.821   &  1.67   & 0.062   \\
				\hline
				BP15 & 19.10 & 72.89 & $2.401\times 10^{-4}$ &  2.62  & 5.348    & 2.329  &   0.074  \\
				\hline
				BP16 & 15.87 & 75.024 & $2.192\times 10^{-4}$ & 2.62   & 4.701    & 2.047 & 0.0718   \\
				\hline	
				BP17 & 15.67 & 78.38 & $2.426\times 10^{-4}$ & 2.64   & 4.603    & 2.02  &  0.063   \\
				\hline					   
				BP18 & 19.76 & 83.14 & $1.662\times 10^{-4}$ &  2.64  & 2.240    & 0.98  & 0.0449  \\
				\hline					    
				BP19 & 20.24 & 76.76 & $1.873\times 10^{-4}$ & 2.62   & 3.740  & \textcolor{blue}{}1.62  &  0.0665   \\
				\hline
				BP20 & 28.15 & 77.04 & $9.39\times 10^{-5}$ &  2.61  & 1.779  &  0.77  &  0.063   \\
				\hline 
				BP21 & 27.085  & 79.40 & $8.134\times 10^{-5}$ & 2.61   & 1.390  & 0.603 &  0.056   \\  
				\hline 
				BP22 & 11.83  & 74.06 & $4.577\times 10^{-4}$ & 2.69   & 10.098   & 4.51  & 0.073   \\ 
				\hline  
				BP23 & 12.285  & 76.51 & $3.377\times 10^{-4}$ & 2.63   & 6.857  &  2.998  &  0.067  \\
				\hline
				BP24 & 13.09  & 75.47 & $3.538\times 10^{-4}$ & 2.65   & 7.526  & 3.31    &  0.0709  \\
				\hline
				BP25 & 14.15  & 74.35 & $3.458\times 10^{-4}$ & 2.62   & 7.554  &  3.29   & 0.072   \\
				\hline 
				BP26 & 11.96  & 78.57 & $3.557\times 10^{-4}$ & 2.69   & 6.644  & 2.97   &   0.062  \\
				\hline
				BP27 & 12.60  & 77.17 & $3.311\times 10^{-4}$ & 2.66    & 6.502  & 2.87  &   0.065  \\
				\hline  
				BP28 & 14.30  & 76.77 & $2.423\times 10^{-4}$ &  2.63  & 5.30  &  2.31   &  0.0729   \\ 
				\hline  
				BP29 & 14.16  & 78.86 & $2.572\times 10^{-4}$ &  2.648  & 4.795  &  2.11  & 0.062    \\
				\hline
				BP30 & 12.91  & 81.94 & $2.004\times 10^{-4}$ &  2.65  & 2.991  & 1.31    &   0.049  \\
				\hline
				BP31 & 16.15 & 81.22 &  $1.843\times 10^{-4}$ & 2.63    & 2.917 &  1.27   & 0.0527	 \\
				\hline
				BP32 & 12.85 &  83.93 & $2.308\times 10^{-4}$ & 2.66  & 2.827 &   1.25    &  0.0408    \\ 
				\hline
				BP33 &  11.63   &  88.72   & $1.325\times 10^{-4}$ & 2.68&  0.830    & 0.369  & 0.0208   \\
				\hline
				BP34 &   19.86   &  88.73   & $8.03\times 10^{-5}$ &  2.61 & 0.502 & 0.0208 & 0.021  \\
				\hline 
				BP35 &  22.71  &  74.16  & $1.093\times 10^{-4}$ &  2.61 & 2.344 & 1.01 & 0.071   \\
				\hline \hline
			\end{tabular}
			|		}		
	\end{center}
	\caption{Extended list of BPs used in the MC simulation for the 2HDM Type-I parameter scan, highlighting the $h$ and $a$ masses as well as the signal LO cross section, QCD $K$-factor and event rate after the full cutflow, together with its significance $\Sigma$  (against the $t\bar t$ background) and efficiency $\epsilon$. Recall that 
NNLO QCD $K$-factors have been used for Higgs production (and the NLO QCD one of $-27\%$ for $ggt\overline{t}$). Here, $\sqrt s=13$ TeV and ${\cal L}=300$ fb$^{-1}$.}
	\label{tab5}
\end{table}

\section{Conclusions}
\label{section5}

In this paper, we have shown the outcome of performing some recasting over the parameter space of the 2HDM Type I, wherein the heaviest CP-even Higgs state $H$  is identified with the discovered SM-like one, $H_{\rm SM}$,  while $h$ and $a$ are lighter. After considering the available experimental data from searches for exotic Higgs decay into two light (pseudo)scalars, we have found that the corresponding parameter space for which there is sensitivity via $H_{\rm SM}\to hh(aa)\to 2b2\tau$ at Run 2 is already excluded by existing constraints from BSM Higgs searches. Furthermore, we have shown that there are regions of the 2HDM Type-I parameter space compliant with theoretical and experimental constraints yielding
substantial BR$(H^\pm \to W^\pm a)$ and BR$(H \to Z^{*}Z^{*}h)$. The large size of the former has been exploited  in other literature. Here, 
 concerning the latter, we have made the case for looking at the process $pp\to H_{\rm SM}\to Z^{*}A\to  Z^{*} Z^{*}h$ in $ \mu^+\mu^- jj b\overline{b}$ final states, specifically,  in the region with large $m_a$ and small $m_h$. After performing a full MC analysis down to detector level, we have proven that the
overwhelming  background arising from top-quark pair production in association with 2 ISR jets can be suppressed after applying efficient kinematics cuts, leading to a large significance of this hitherto unexplored light Higgs signature already at Run 3 of the LHC, where evidence of it can be seen, further affording one with clear discovery potential at the HL-LHC.   

\section*{Acknowledgements}
The work of SM is supported in part through the NExT Institute and the STFC Consolidated Grant No. ST/L000296/1. SS is fully supported through the NExT Institute.


\appendix

\end{document}